\documentclass{article} 
\usepackage{iclr2025_conference,times}

\usepackage{amsmath,amsfonts,bm}

\def\eqref#1{equation~\ref{#1}}

\def\1{\bm{1}}

\DeclareMathAlphabet{\mathsfit}{\encodingdefault}{\sfdefault}{m}{sl}
\SetMathAlphabet{\mathsfit}{bold}{\encodingdefault}{\sfdefault}{bx}{n}

\usepackage{hyperref}
\usepackage{url}
\usepackage{booktabs}
\usepackage{makecell}
\usepackage{pifont}
\usepackage{amssymb}
\usepackage{tikz}
\usepackage{booktabs,tabularx,hyperref}
\usepackage{framed}

\definecolor{Gray}{gray}{0.91}
\definecolor{LightCyan}{rgb}{0.82,0.82,1}
\newcolumntype{a}{>{\columncolor{Gray}}c}
\newcolumntype{B}{>{\columncolor{LightCyan}}c}

\definecolor{blue}{HTML}{a4c3e3}
\definecolor{pink}{HTML}{8edbc2}
\definecolor{FIRST}{HTML}{d4e9d8}
\definecolor{SECOND}{HTML}{fdf3d0}

\usepackage{pifont}
\newcommand{\cmark}{\ding{51}}     
\newcommand{\xmark}{\ding{55}}     
\newcommand{\nrmark}{\textsc{n/r}} 

\usepackage[T1]{fontenc}
\usepackage{textcomp}
\usepackage{multirow}
\usepackage{wrapfig}
\usetikzlibrary{mindmap, shadows}
\usepackage[HTML]{xcolor}
\definecolor{darkgreen}{HTML}{006400}
\usepackage{caption}
\usepackage{soul} 
\usepackage{xcolor}

\usepackage[table]{xcolor}
\usepackage{footmisc}  
\usepackage[most]{tcolorbox}
\usepackage{tcolorbox}
\tcbuselibrary{minted,breakable,skins}
\usepackage{enumitem}

\lstdefinelanguage{JSON}{
  morestring=[b]",
  morecomment=[l]{//},
  literate=
   *{0}{{{\char`0}}}{1}
    {1}{{{\char`1}}}{1}
    {2}{{{\char`2}}}{1}
    {3}{{{\char`3}}}{1}
    {4}{{{\char`4}}}{1}
    {5}{{{\char`5}}}{1}
    {6}{{{\char`6}}}{1}
    {7}{{{\char`7}}}{1}
    {8}{{{\char`8}}}{1}
    {9}{{{\char`9}}}{1}
}

\newtcolorbox{PromptBox}[1][]{
  breakable,
  enhanced,
  colback=gray!3,
  colframe=gray!50,
  colbacktitle=gray!15,
  coltitle=black,         
  fonttitle=\bfseries,    
  title={#1},
  left=1em, right=1em, top=0.8em, bottom=0.8em,
  boxrule=0.5pt, arc=2pt
}

\newcommand{\red}[1]{{#1}}

\usetikzlibrary{arrows.meta, positioning, fit, shapes.multipart}

\usepackage{algorithm}
\usepackage{algpseudocode}

\usepackage{booktabs,tabularx,array}
\usepackage{xurl}         
\usepackage{hyperref}
\hypersetup{hidelinks,breaklinks=true}

\newcolumntype{Y}{>{\raggedright\arraybackslash}X}

\usepackage{etoolbox}
\usepackage{risys-title}

\RisysTitle{RedSage: A Cybersecurity Generalist LLM}
\RisysAuthor[1]{Naufal Suryanto}
\RisysAuthor[1,*]{Muzammal Naseer}
\RisysAuthor[1]{Pengfei Li}
\RisysAuthor[2]{Syed Talal Wasim}
\RisysAuthor[2]{Jinhui Yi}
\RisysAuthor[2]{Juergen Gall}
\RisysAuthor[3]{Paolo Ceravolo}
\RisysAuthor[3]{Ernesto Damiani}
\RisysAffil[1]{Khalifa University}
\RisysAffil[2]{University of Bonn}
\RisysAffil[3]{University of Milan}
\RisysContribution[*]{Project lead}
\RisysProjectpage{https://risys-lab.github.io/RedSage/}

\RisysAbstract{Cybersecurity operations demand assistant LLMs that support diverse workflows without exposing sensitive data. Existing solutions either rely on proprietary APIs with privacy risks or on open models lacking domain adaptation. To bridge this gap, we curate 11.8B tokens of cybersecurity-focused continual pretraining data via large-scale web filtering and manual collection of high-quality resources, spanning 28.6K documents across frameworks, offensive techniques, and security tools.
Building on this, we design an agentic augmentation pipeline that simulates expert workflows to generate 266K multi-turn cybersecurity samples for supervised fine-tuning. Combined with general open-source LLM data, these resources enable the training of RedSage, an open-source, locally deployable cybersecurity assistant with domain-aware pretraining and post-training.
To rigorously evaluate the models, we introduce RedSage-Bench, a benchmark with 30K multiple-choice and 240 open-ended Q\&A items covering cybersecurity knowledge, skills, and tool expertise. 
RedSage is further evaluated on established cybersecurity benchmarks (e.g., CTI-Bench, CyberMetric, SECURE) and general LLM benchmarks to assess broader generalization. At the 8B scale, RedSage achieves consistently better results, surpassing the baseline models by up to +5.59 points on cybersecurity benchmarks and +5.05 points on Open LLM Leaderboard tasks. These findings demonstrate that domain-aware agentic augmentation and pre/post-training can not only enhance cybersecurity-specific expertise but also help to improve general reasoning and instruction-following. All models, datasets, and code are publicly available.}

\iclrfinalcopy 

\begin{document}

\RisysMakeTitle




\begin{figure}[!b]
    \centering
    \includegraphics[width=1\linewidth]{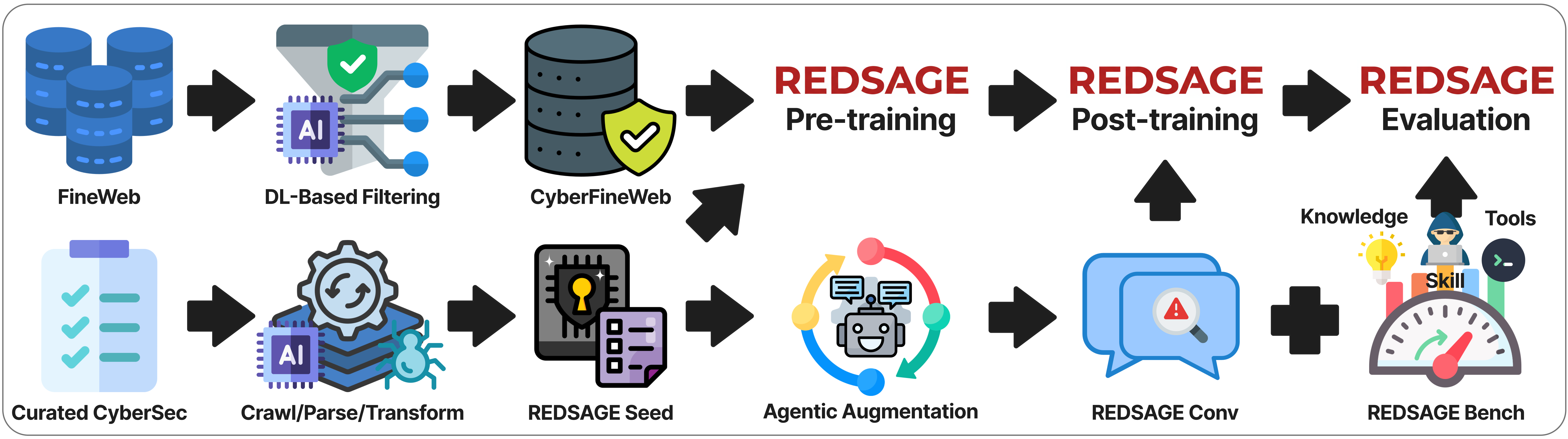}
    \caption{Overview of the RedSage pipeline. RedSage is trained through continual pre-training on cybersecurity-filtered corpora and post-training with curated and augmented conversation data, followed by evaluation on a comprehensive benchmark covering knowledge, skills, and tool expertise.}
    \label{fig:redsage-overview}
\end{figure}

\section{Introduction}
The rapid evolution of cybersecurity threats has elevated the need for proactive and comprehensive defense strategies, as organizations face increasingly sophisticated attacks and advanced persistent threats \citep{2024systematicliterature}. 
Modern cybersecurity involves a wide range of critical tasks, including threat analysis, incident response, vulnerability management, and security monitoring. 
However, the complexity of security tools and the level of expertise required to operate them present significant challenges. 
These challenges are compounded by a global skills shortage, with research estimating a demand–supply gap of millions of unfilled cybersecurity positions \citep{isc22024workforce}. 
Consequently, there is growing momentum to employ cybersecurity-tuned LLMs to augment human analysts.

Recent efforts have produced cybersecurity-trained LLMs, yet most emphasize a single training stage while overlooking others. For instance, some extend pretraining on domain-specific corpora \citep{kassianik2025foundationsec} but apply limited post-training with only 835 samples \citep{yu2025primus} or fewer than 30K cybersecurity-filtered items \citep{weerawardhena2025foundationsec_instruct}, while others focus on supervised fine-tuning with large cybersecurity Q\&A collections without pretraining to strengthen domain knowledge \citep{deephat2025}. Further, existing cybersecurity benchmarks offer only partial coverage, such as omitting tool proficiency and qualitative evaluation of free-response Q\&A beyond simple MCQs (see Table~\ref{tab:benchmarks_comparison} and Fig.~\ref{fig:redsage_taxonomy}). Beyond these gaps, most works also do not release their data or pipelines, limiting reproducibility and openness (see Table~\ref{tab:training-data-features}).


To address these gaps, we present RedSage (\textbf{R}etrieval-\textbf{E}nhanced \textbf{D}ata-driven \textbf{S}ecurity \textbf{A}ssistant \textbf{G}uidance and \textbf{E}valuation), an open-source LLM tailored for cybersecurity. As illustrated in Fig.~\ref{fig:redsage-overview}, RedSage integrates large-scale continual pretraining on cybersecurity-filtered corpora, post-training with curated and agentically augmented datasets, and rigorous evaluation across knowledge, skills, and tool proficiency. Our key contributions are:  
(1) assembling an 11.8B-token corpus of cybersecurity data for domain-specific continual pretraining,  
(2) constructing a 266K-sample augmented dataset via an agentic pipeline for supervised fine-tuning, followed by preference alignment with open-source data,
(3) introducing RedSage-Bench, a benchmark with 30K MCQs for broad coverage and 240 open-ended Q\&A items for quality evaluation across knowledge, skills, and tools, and 
(4) RedSage, an open 8B model with data and code, achieving state-of-the-art results on established cybersecurity benchmarks while also improving on general benchmarks.   

\begin{table}[!t]
\centering
\begin{minipage}{0.65\textwidth}
\centering\small
\caption{
Comparison of cybersecurity LLM benchmarks. Columns indicate knowledge (Know.), skills (Skill), tool proficiency (Tool), and use of quality scoring (Qual.). Size = total samples. Agentic CTF benchmarks (e.g., NYU-CTF, CyBench) are excluded as they are interactive rather than base LLM eval.
}
\label{tab:benchmarks_comparison}
\setlength{\tabcolsep}{6pt}
		\scalebox{0.950}[0.950]{
\begin{tabular}{lccccr}
\toprule
\rowcolor{Gray} 
\textbf{Name} & \textbf{Know.} & \textbf{Skill} & \textbf{Tool} & \textbf{Qual.} & \textbf{Size} \\
\midrule
SecEval       & \checkmark & \xmark     & \xmark     & \xmark & 2,000  \\
CyberMetric   & \checkmark & \xmark     & \xmark     & \xmark & 10,000 \\
CyberBench    & \checkmark & \xmark     & \xmark     & \xmark & 80,422 \\
SECURE        & \checkmark & \xmark     & \xmark     & \xmark & 4,072  \\
CS-Eval       & \checkmark & \xmark     & \xmark     & \xmark & 4,369  \\
SecBench      & \checkmark & \xmark     & \xmark     & \xmark & 47,910 \\
CTI-Bench     & \checkmark & \checkmark & \xmark     & \xmark & 5,610  \\
CyberSecEval  & \xmark     & \checkmark & \xmark     & \xmark & 1,000  \\
\hline
RedSage-Bench (Ours)          & \checkmark & \checkmark & \checkmark & \checkmark & 30,240 \\
\bottomrule
\end{tabular}}

\end{minipage}%
\hfill
\begin{minipage}{0.32\textwidth}
\centering \small 
\includegraphics[width=\linewidth]{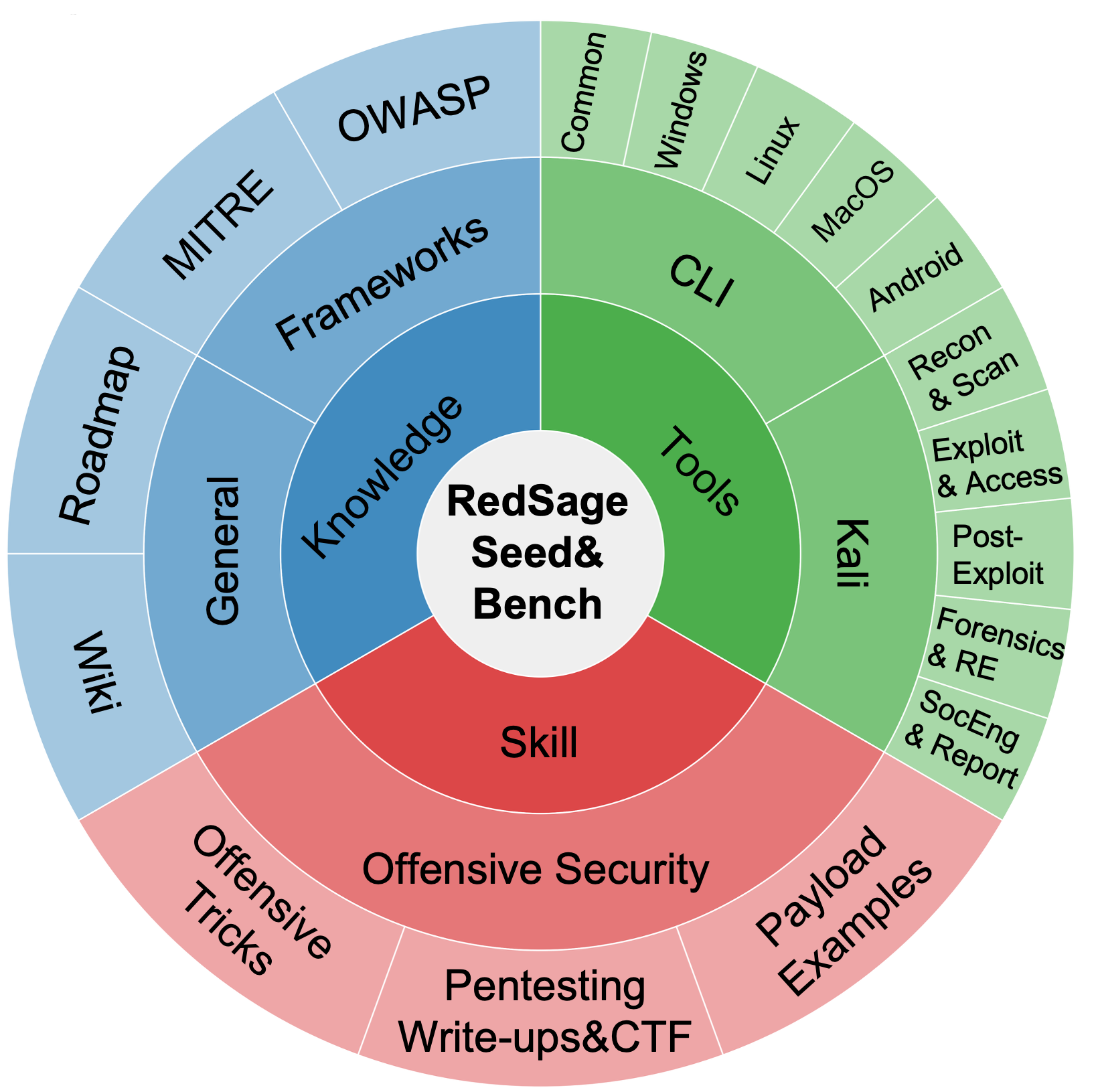}
\captionof{figure}{
Taxonomy of RedSage Seed\&Bench dataset. 
It spans knowledge, practical offensive skills, 
and tool expertise (CLI and Kali Linux).
}
\label{fig:redsage_taxonomy}
\end{minipage}
\end{table}

\section{Related Works}

\subsection{Cybersecurity Benchmarks}

\textbf{General Knowledge.} 
Several benchmarks assess LLMs’ understanding of core cybersecurity concepts via structured Q\&A. 
\textit{SecEval}~\citep{li2023seceval} includes 2K+ MCQs across nine domains (web, system, application security). 
\textit{CyberMetric}~\citep{cybermetric24} provides 10K MCQs generated with RAG and expert validation, spanning penetration testing and network security. 
\textit{CyberBench}~\citep{liu2024cyberbench} extends beyond MCQs to tasks such as NER, summarization, and classification. 
\textit{SECURE}~\citep{secure2024} targets Industrial Control Systems with domain-specific MCQs on risk reasoning and vulnerability analysis. 
\textit{CS-Eval}~\citep{yu2024cseval} covers 42 subcategories across three cognitive levels (Knowledge, Ability, Application) using MCQs, multi-answer, T/F, and open-ended items. 
\textit{SecBench}~\citep{jing2024secbench} offers 44,823 MCQs and 3,087 SAQs in Chinese and English, capturing both factual recall and logical reasoning.  

\textbf{Applications and Agentic Tasks.} 
Application-oriented benchmarks probe reasoning beyond recall. 
\textit{CTIBench}~\citep{ctibench2024} defines four tasks: MCQs, common vulnerabilities and exposures(CVE)-to-common weakness enumeration (CWE) mapping, common vulnerability scoring system (CVSS) prediction, and threat actor attribution in cyber threat intelligence. 
\textit{CyberSecEval}~\citep{wan2024cyberseceval} examines model risks across eight areas (e.g., exploit generation, prompt injection). 
Agentic evaluations such as \textit{NYU-CTF}~\citep{shao2024nyu} and \textit{CyBench}~\citep{zhang2025cybench} assess red-team capabilities through capture-the-flag (CTF) challenges in interactive settings.  

While these efforts advance evaluation of knowledge and applications, they rarely isolate competence in understanding and operating security tools or systematically assess the quality of free-form responses. As summarized in Table~\ref{tab:benchmarks_comparison}, most benchmarks specialize in either knowledge or applications, and even agentic ones lack explicit tool-focused assessment. We address these gaps with RedSage-Bench, which jointly measures knowledge, skills, and tool proficiency (Fig.~\ref{fig:redsage_taxonomy}).

\subsection{Cybersecurity Datasets and Models}

\textbf{Early Cybersecurity Datasets.} Early domain-specific models such as \textit{CyBERT}~\citep{ranade2021cybert}, \textit{SecureBERT}~\citep{aghaei2022securebert}, and \textit{CTI-BERT}~\citep{park2023pretrained} showed the value of domain-adaptive fine-tuning. However, their datasets were not released and were task-specific.

\textbf{Cybersecurity Datasets for LLMs.}  With the advent of LLMs, several groups curated cybersecurity-specific corpora. 
\textit{PRIMUS}~\citep{yu2025primus} (Trend Micro) provides 2.75B tokens for continued pretraining, 835 samples for supervised fine-tuning, and reasoning distillation, extending Llama-3.1-8B-Instruct into Llama-Primus-Base and -Merged. 
\textit{Foundation-Sec-8B}~\citep{kassianik2025foundationsec} (Cisco) collects 5.1B tokens via large-scale scraping and filtering, continues pretraining on Llama-3.1-8B-Base, and adds a cybersecurity post-training stage, though its dataset remains closed. 
Community efforts include \textit{DeepHat} (formerly WhiteRabbitNeo), reportedly trained on 1M+ Q\&A pairs for real workflows~\citep{deephat2025}, and \textit{Lily-Cybersecurity}, which fine-tunes Mistral-7B on 22K hand-crafted and lightly refined conversations~\citep{lily2024}. 
\textit{Cyber-DAP}~\citep{salahuddin2025lessdata} highlights the effectiveness of smaller curated corpora for continued pretraining, while \textit{SecGemini}~\citep{secgemini2025} offers a closed model with live threat intelligence but unreleased data. We summarize these datasets in Table~\ref{tab:training-data-features}.

\begin{table}[!t]
\centering
\caption{ \small
 Comparison of cybersecurity-tuned LLM training datasets.
Pretraining and curated columns report token counts (B = billion, M = million). SFT reports the number of supervision samples. \cmark = present; \xmark = absent; \nrmark = not reported.
}
\vspace{-1em}
\label{tab:training-data-features}
\setlength{\tabcolsep}{11pt}
		\scalebox{0.850}[0.850]{
\begin{tabular}{lcccccc}
\toprule
\rowcolor{Gray} 
\textbf{Name} & \makecell{Pretrain\\Tokens (B)} & \makecell{Curated\\Tokens (M)} & \makecell{SFT\\Samples} & \makecell{Agentic\\Augmented} & \makecell{Open\\Data} & \makecell{Open\\Model} \\
\midrule
PRIMUS                          & 2.57 & 191   & 835             & \xmark & \cmark & \cmark \\
Foundation\mbox{-}Sec\mbox{-}8B & 5.10 & \xmark & 28K             & \xmark & \xmark & \cmark \\
DeepHat             & \xmark & \xmark & $>$1M           & \xmark & \xmark & \cmark \\
Lily\mbox{-}Cybersecurity\mbox{-}7B & \xmark & \xmark & 22K             & \xmark & \xmark & \cmark \\
Cyber\mbox{-}DAP                 & \xmark & 119   & \xmark          & \xmark & \xmark & \xmark \\
SecGemini (closed)              & \nrmark & \nrmark & \nrmark       & \xmark & \xmark & \xmark \\
\textbf{Ours (RedSage)} & \textbf{11.7} & \textbf{850} & \textbf{266K} & \cmark & \cmark & \cmark \\
\bottomrule
\end{tabular}}

 \small
 Dataset statistics are compiled from official publications, technical reports, and model cards.
\vspace{-1.5em}
\end{table}


Unlike prior work with limited augmentation, we introduce \emph{agentic augmentation} to transform curated cybersecurity resources into diverse, realistic multi-turn dialogs simulating expert–assistant workflows across knowledge, offensive operations, and tool proficiency for domain-specific fine-tuning.
RedSage is, to our knowledge, the only effort combining large-scale continual pretraining, curated data, agentically augmented SFT, and full openness (data, model, and code) (Table~{\ref{tab:training-data-features}}).

\section{RedSage}
\label{section:proposal}

\begin{figure}[!t]
    \centering
    \includegraphics[width=\linewidth]{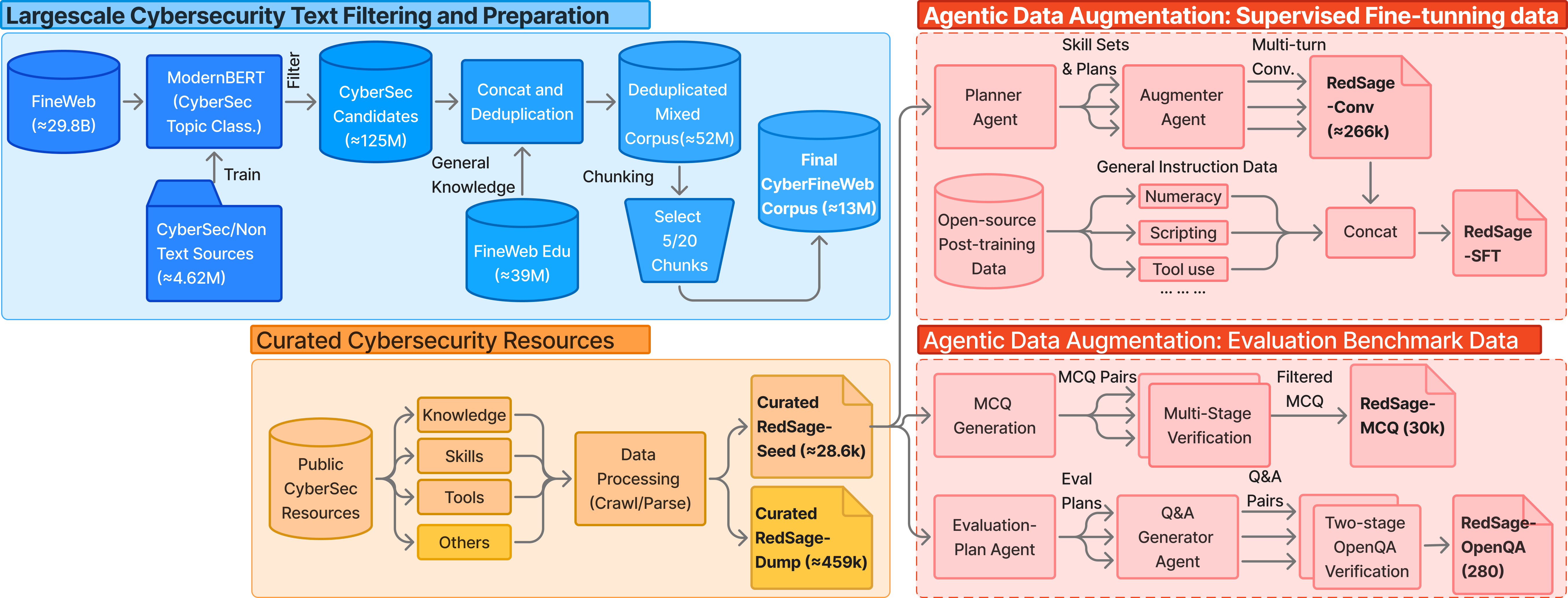}
    \caption{\small
        RedSage data pipeline combining large-scale text collection, curated cybersecurity resources, 
        and agentic augmentation for supervised fine-tuning and benchmark generation. \emph{Best viewed in Zoom.}
    }
    \vspace{-1em}
    \label{fig:redsage-pipeline}
\end{figure}

We build RedSage through a data-centric pipeline comprising  
(1) large-scale filtering of cybersecurity text and curation of high-quality resources for continual pretraining,  
(2) agentic augmentation to create supervised fine-tuning data, and  
(3) benchmark construction for evaluation (Fig.~\ref{fig:redsage-pipeline}).


\subsection{RedSage Pre-training Data}
\label{subsec:redsage-pretraining-data}


\textbf{CyberFineWeb.}
We construct CyberFineWeb by filtering FineWeb~\citep{penedo2024the}, a cleaned large-scale web corpus aggregated from Common Crawl (2013--2024; $\sim$15T tokens). 
To extract cybersecurity content, we fine-tune a binary classification model based on ModernBERT-base~\citep{warner2024smarter}, a state-of-the-art encoder trained on 2T+ tokens. 
Applying this filter yields a \emph{cybersecurity candidate pool} of $\sim$125M documents ($\sim$89.8B tokens).

To avoid catastrophic forgetting on general knowledge, we mix CyberFineWeb with general-knowledge samples from FineWeb-Edu~\citep{lozhkov2024fineweb-edu} at a 30\% replay ratio. 
FineWeb-Edu is a 1.3T-token educational subset shown to improve general LLM benchmarks. 
This strategy follows prior work on replay-based continual learning~\citep{ibrahim2024simple,guo2025efficient}, though unlike dynamic replay, we embed these examples directly into the static corpus. 
We then apply global near-duplicate removal with MinHash-LSH over the combined data. 
This yields a deduplicated mixed corpus of $\sim$52M documents ($\sim$46.8B tokens), while inheriting FineWeb’s upstream extensive filtering and PII removal.

Finally, we partition the deduplicated corpus into 20 chronological chunks for sequential training under compute constraints and apply early stopping after 5 chunks to control training cost. This yields the \emph{final CyberFineWeb corpus}: $\sim$13M documents ($\sim$11.7B tokens) used in our model.
Implementation details, including classifier training, 
deduplication parameters, and datasets statistics, are provided in Appendix~\ref{app:cyberfineweb}.

\textbf{RedSage-Seed.}
Web-filtered text offers broad coverage, but its reliability is not assured. 
To provide high-quality content, we curate RedSage-Seed: 28{,}637 samples ($\sim$0.15B tokens) from publicly available sources organized into three categories: 
\emph{Knowledge} (well-established cybersecurity frameworks and knowledge bases~\citep{mitre_attack,mitre_capec,mitre_cwe,owasp_top10}), 
\emph{Skills} (penetration-testing write-ups~\citep{0df}, hacking techniques~\citep{hacktricks}, payload examples~\citep{payloads_all_the_things}, and ethical hacking tutorials/blogs~\citep{nullbyte,hackingarticles}), and 
\emph{Tools} (CLI cheat-sheets~\citep{tldr_pages}, Linux manuals~\citep{linux_man_pages}, Kali tools~\citep{kali_tools}). 
We additionally collect an uncategorized dump of $\sim$459K documents ($\sim$0.7B tokens) from trusted cybersecurity sources (Appendix~\ref{app:redsageseed}) to supply extra pretraining tokens.


To process these resources, we crawl web-based sources and convert them to Markdown using ReaderLM-v2~\citep{wang2025readerlmv2}, while downloadable resources are parsed directly. 
This hierarchical Markdown format preserves structure and enables effective chunking for  subsequent augmentation stages. Only the categorized seeds are used for augmentation, while both sets support pretraining. Full statistics, categorization, processing steps, and examples are in Appendix~\ref{app:redsageseed}.


\begin{table*}[ht]
\centering \small 
\caption{
Statistics of RedSage-Seed (curated pretraining corpus) vs.\ RedSage-Conv (augmented SFT data) by category. Columns show sample counts, average tokens, and total tokens.
}
\label{tab:redsage_seed_conv_comparison}
\setlength{\tabcolsep}{7pt}
		\scalebox{0.90}[0.90]{
\begin{tabular}{lrrrrrr}
\toprule
\rowcolor{Gray} 
\textbf{Category}
& \multicolumn{3}{c}{\textbf{Seed}} 
& \multicolumn{3}{c}{\textbf{Conversation}} \\
\cmidrule(lr){2-4} \cmidrule(lr){5-7}
& \textbf{Samples} & \textbf{Avg. Tokens} & \textbf{Tokens (M)} 
& \textbf{Samples} & \textbf{Avg. Tokens} & \textbf{Tokens (M)} \\
\midrule
Knowledge – General     & 6,924  & 2,370 & 16.4  & 67,635  & 1,326 & 89.6 \\
Knowledge – Frameworks  & 3,715  & 2,935 & 10.5  & 39,908  & 1,285 & 51.0 \\
Skill – Offensive       & 4,032  & 9,478 & 37.8  & 38,870  & 1,345 & 52.3 \\
Tools – CLI             & 12,943 & 5,774 & 78.9  & 109,261 & 1,331 & 145.7 \\
Tools – Kali            & 1,023  & 6,693 & 6.3   & 10,506  & 1,356 & 14.3 \\
\midrule
\textbf{Total}          & \textbf{28,637} & \textbf{5,231} & \textbf{149.8} 
                        & \textbf{266,180} & \textbf{1,326} & \textbf{353.0} \\
\midrule
Cybersecurity Dumps     & 459,473   & 1,524 & 700.1   & --  & -- & -- \\
\bottomrule
\end{tabular}}
\end{table*}

\subsection{RedSage Post-training Data}
\label{subsubsec:redsageconv}

\begin{figure}
    \centering
    \includegraphics[width=1\linewidth]{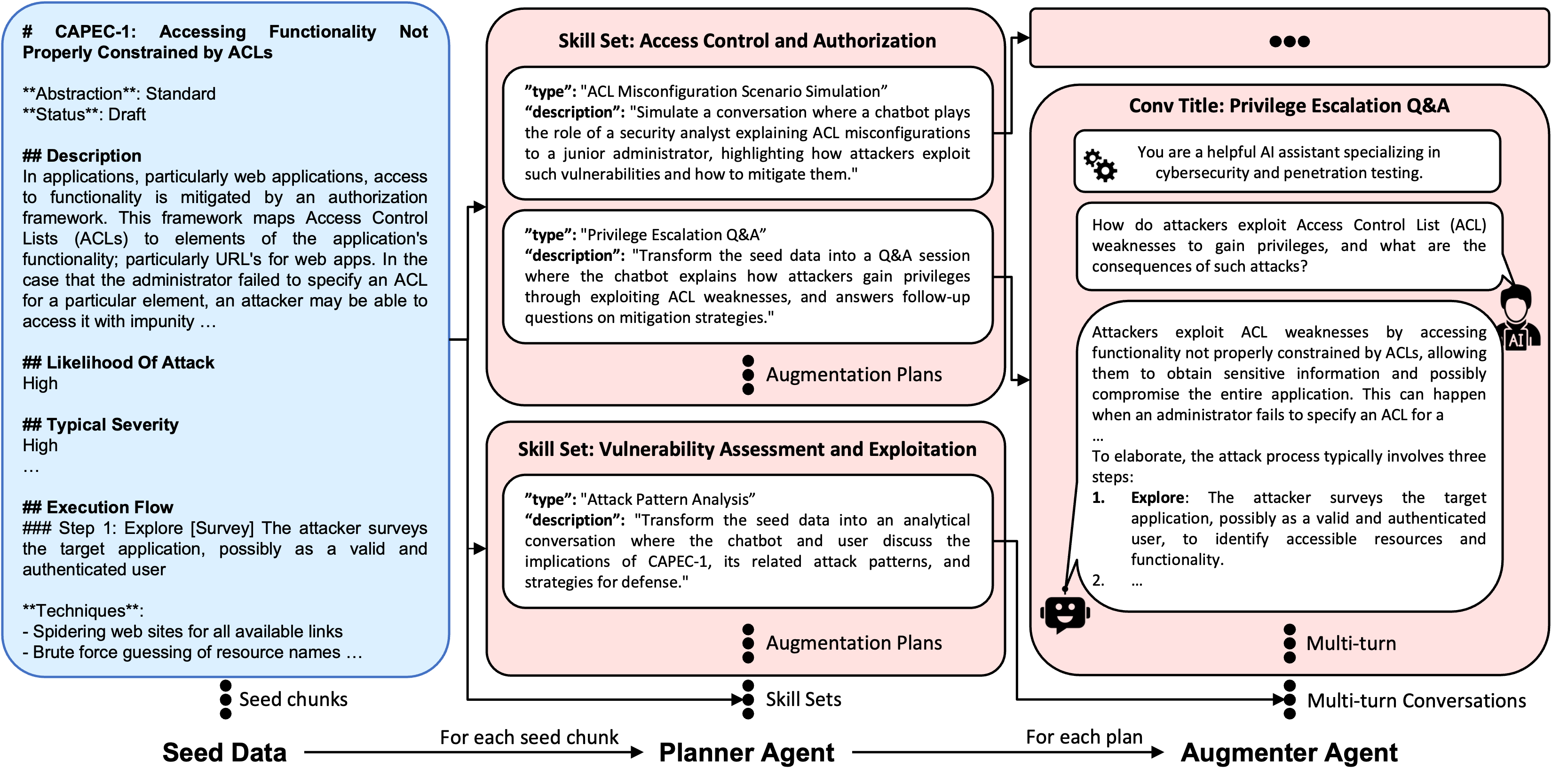}
    \caption{ \small
    Agentic data augmentation pipeline. 
    Seed data (e.g., CAPEC attack patterns) is processed by the \emph{Planner Agent} into skill sets and augmentation plans, which the \emph{Augmenter Agent} instantiates as grounded, role-based multi-turn cybersecurity dialogues for supervised fine-tuning (SFT).
    }
    \label{fig:redsage-agentic}
\end{figure}

\textbf{Agentic Data Augmentation.}
To enable assistants capable of realistic security dialogues, we augment RedSage-Seed into multi-turn conversations using an agentic framework inspired by AgentInstruct~\citep{mitra2024agentinstruct}.
Unlike prior work with fixed skill templates, our \emph{Planner Agent} analyzes each seed data chunk and derives candidate skill sets (e.g., vulnerability analysis, tool-command generation) along with augmentation strategies that describe how the seed is transformed, adapted into a conversational or Q\&A format, and enriched with explanations. We enforce guidelines on relevance, diversity, creativity, detail, and formatting.
The \emph{Augmenter Agent} then instantiates each plan into realistic, role-based multi-turn dialogues grounded in the seed data.
This pipeline scales efficiently, producing multiple dialogues per skill set and filtering outputs for format validity, consistency, and topical relevance. 
Overall, it yields RedSage-Conv with $\sim$266K multi-turn conversations ($\sim$352M tokens), expanding total samples by 9.2$\times$ and tokens by 2.3$\times$ across knowledge, skills, and tools while preserving technical depth (Tab.~\ref{tab:redsage_seed_conv_comparison}). 
Fig.~\ref{fig:redsage-agentic} illustrates the augmentation pipeline, while detailed statistics, prompts, and examples are provided in Appendix~\ref{app:redsageconv}.

\textbf{General instruction integration.}
While domain-specific conversations ground the assistant in cybersecurity, effective models must also handle broader instruction-following tasks. 
We therefore complement RedSage-Conv with curated post-training SFT data from SmolLM3~\citep{bakouch2025smollm3}\footnote{General SFT datasets: \href{https://huggingface.co/HuggingFaceTB}{HuggingFaceTB/smoltalk2}}, focusing on its non-reasoning subset. 
This corpus adds coverage of summarization, numeracy, data interpretation, temporal and unit reasoning, commonsense knowledge, step-by-step planning, technical writing, scripting, and general tool use. 
The combination of cybersecurity-specific and general instruction data yields a high-quality post-training corpus, enabling a cybersecurity assistant that performs specialized tasks while retaining broad capabilities.

\subsection{RedSage Benchmark}
\label{subsubsec:redsagebench}

\textbf{Multiple-choice Q\&A generation.}
We derive MCQs from RedSage-Seed as follows: for each seed item, a strong open instruction-tuned LLM 
\footnote{Teacher and Verifier LLM: \label{fn:teacherllm}\href{https://huggingface.co/meta-llama/Llama-3.3-70B-Instruct}{meta-llama/Llama-3.3-70B-Instruct}, \href{https://huggingface.co/Qwen/Qwen2.5-72B-Instruct}{Qwen/Qwen2.5-72B-Instruct}}
generates several MCQs under guidelines: items are self-contained and closed-book, target stable domain facts/procedures, follow a four-option format with three plausible distractors, and satisfy diversity and formatting constraints.

\textbf{Open-ended Q\&A generation.}
We extend RedSage-Seed into open-ended Q\&A using an agentic augmentation framework with two stages: (1) an \emph{Evaluation-Planner} analyzes seed artifacts and proposes realistic evaluation types with instruction templates and answer guidelines; (2) a \emph{Question-Answer Generator} instantiates each plan into a self-contained open-ended Q\&A  with a natural-language prompt and a reference answer. 
All open-ended Q\&A  are grounded in the seed data and scored with a reference-based LLM-as-judge rubric that evaluates both factual correctness (True/False) and answer quality (0–10) across helpfulness, relevance, depth, and level of detail.

\paragraph{Multi-stage verification.}
For MCQs, we apply a two-stage pipeline: \emph{Stage~1 (structural validity)} uses a verifier LLM 
\footref{fn:teacherllm}
with a checklist on format, correctness, distractors, topical relevance, and consistency, filtering items by pass/fail; \emph{Stage~2 (quality scoring)} then applies the same verifier LLM \footref{fn:teacherllm} to
assign each remaining item a score \(s\in[0,10]\) for clarity, correctness, and assessment value. 
\red{In both stages, we use chain-of-thought prompting so the verifier explicitly reasons through each checklist criterion before issuing a pass/fail label or score, yielding judgments that more closely follow our rubric.}
We then select the pairs where $s > 8$
and apply quota-aware random sampling
to ensure taxonomic balance, yielding \(30{,}000\) MCQ–answer pairs evenly split across knowledge, skills, and tools. For open-ended Q\&A, we directly perform LLM-based quality scoring in \emph{Stage~2}
followed by human verification, selecting \(240\) high-quality pairs evenly distributed across categories. 

\paragraph{\red{Human quality control.}}
\red{Across all verification stages, we iteratively refined prompts and manually inspected sampled outputs until the verifier consistently aligned with our criteria. We observe that chain-of-thought prompting plays a significant role in producing more precise judgments. For the large-scale MCQ benchmark, random audits confirmed that items passing the final stages met both Stage~1 and Stage~2 requirements. For open-ended Q\&A, we retain only human-verified items.}

\paragraph{\red{Data decontamination.}}
\red{We apply an additional filtering and deduplication step to prevent unintended overlap between our benchmark datasets and augmented post-training data, despite their being generated through different pipelines and output formats. Specifically, we remove any synthetic post-training instance whose query has a semantic similarity above 0.9 to a benchmark question. This eliminates 2.96\% of data relative to the benchmark size (0.31\% of the full training corpus) and helps ensure that evaluation remains free of training leakage.}


Implementation details, intermediate outputs, prompt templates, and qualitative examples are provided in Appendix~\ref{app:redsage-bench}\red{, and the full evaluation protocol is described in Appendix~\ref{app:redsage-benchmarks}.}


\subsection{RedSage Training}
\label{subsection:model_training}

\begin{figure}[t]
    \centering
    \vspace{-3mm}
    \includegraphics[width=1\linewidth]{figures/RedSage-Training-Pipeline.pdf}
    \vspace{-3mm}
    \caption{ \small
    \red{RedSage training pipeline. We first continue pretraining the Qwen3 base model on CyberFineWeb to obtain RedSage-CFW, followed by RedSage-Seed and RedSage-Dump to produce RedSage-Base. We then perform supervised fine-tuning using RedSage-Conv and SmolTalk2 \citep{bakouch2025smollm3} data, and finalize the model with Direct Preference Optimization using the Tulu3 Preference Mixture \citep{lambert2025tulu3pushingfrontiers}.}
    }
    \label{fig:redsage-training-pipeline}
\end{figure}

We build RedSage using the Axolotl framework~\citep{axolotl}, with continued pretraining of the open-source base model, Qwen3-8B-Base~\citep{yang2025qwen3},  on cybersecurity corpora, followed by post-training through supervised fine-tuning on augmented conversations and preference alignment. \red{We illustrate training stages in Fig.~\ref{fig:redsage-training-pipeline} with further training details, including exact hyperparameters, estimated training time, and computational cost analysis in Appendix \ref{app:training-details}.}




\textbf{Training setup.}
For continued pretraining (CPT), we first train on the CyberFineWeb corpus, followed by training on RedSage-Seed (Sec.~\ref{subsec:redsage-pretraining-data}). 
We run a single epoch with distributed optimization on 32$\times$A100-64GB GPUs (global batch size 1024), using DeepSpeed ZeRO Stage~3, the AdamW optimizer, and a fixed learning rate of \(2.5\times10^{-6}\) with linear warmup.
After pre-training, we further fine-tune our base model on RedSage-Conv and general SFT data (Sec.~\ref{subsubsec:redsageconv}) with two epochs using a cosine learning rate schedule.
%
We apply direct preference optimization (DPO)~\citep{rafailov2023direct} with open-source Tulu~3 8B Preference Mixture dataset~\citep{lambert2025tulu3pushingfrontiers} 
using original hyperparameters.

\section{Experiments and Results}
\label{section:experiments}

We evaluate the performance of our cybersecurity-tuned LLM on (1) our own benchmark (Sec.~\ref{subsubsec:redsagebench}), (2) related cybersecurity benchmarks, and (3) general LLM benchmarks. 



\textbf{Evaluation setup.}
For replicable results, we implement and evaluate RedSage-Bench and prior cybersecurity benchmarks in HuggingFace \texttt{lighteval}~\citep{lighteval}. 
MCQ benchmarks are scored with normalized log-likelihood accuracy over answer options, while instruction-tuned models and structured output tasks use prefix exact match or regex matching on greedy decoding outputs (temperature=0). Details for each task are provided in Appendix~\ref{app:evaluation-setups}.

\textbf{Baseline methods.}
We evaluate RedSage against both open general-purpose and cybersecurity-tuned LLMs. General-purpose baselines include Llama-3.1-8B~\citep{llama3} and Qwen3-8B~\citep{yang2025qwen3}, while specialized baselines include Llama-Primus (Base, Merged)~\citep{yu2025primus}, Foundation-Sec (Base, Ins)~\citep{kassianik2025foundationsec,weerawardhena2025foundationsec_instruct}, Lily-Cybersecurity-7B-v0.2~\citep{lily2024}, and DeepHat-V1-7B~\citep{deephat2025}. 
\red{We also include Qwen3-32B and GPT-5 \citep{openai2025gpt5systemcard} to compare against larger-capacity and proprietary general-purpose models.}
Base models are evaluated with text completion, instruction-tuned ones with official prompt templates, and we ran \red{hybrid model} in non-reasoning mode for fairness.




Our RedSage variants include three base models: \textbf{RedSage-8B-CFW} (CyberFineWeb only), \textbf{RedSage-8B-Seed} (Seed only), and \textbf{RedSage-8B-Base} (CyberFineWeb followed by Seed). We further derive instruction-tuned variants: \textbf{RedSage-8B-Ins} (instruction-tuned from Base) and the final \textbf{RedSage-8B-DPO}, which combines all data and applies DPO alignment \red{(see Fig.~\ref{fig:redsage-training-pipeline})}.
\red{An additional larger-model scaling experiment is presented in Appendix~\ref{app:larger-scaling}, where partial RedSage data improves a Qwen3-32B model via lightweight QLoRA fine-tuning, demonstrating that our curation pipeline transfers effectively to higher-capacity LLMs.}

\subsection{Evaluation Results on RedSage-Bench}
\label{subsec:redsage-mcq}

\begin{figure}
    \centering
    \includegraphics[width=1\linewidth]{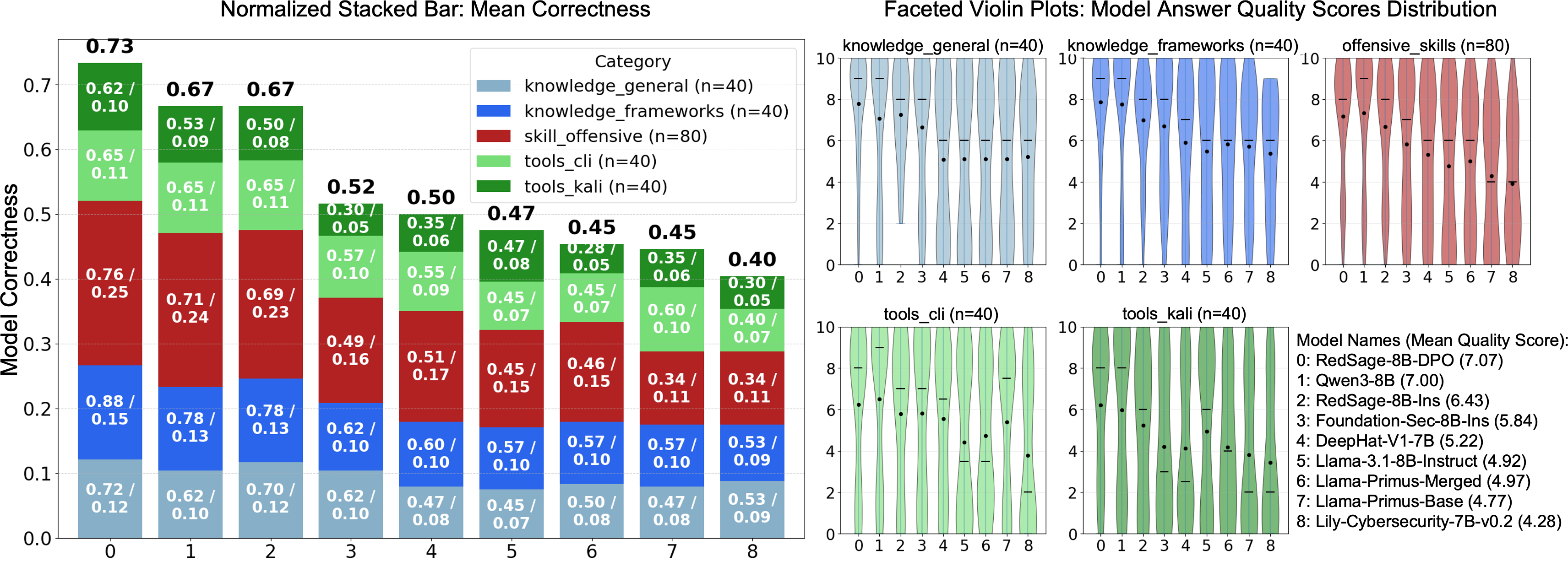}
    \caption{\small 
    \textbf{RedSage open-ended QA evaluation}. Left: normalized stacked bar charts of mean correctness by category (0-1); segment labels show the mean and its relative contribution. Models are ordered by overall mean correctness.
    Right: faceted violin plots of LLM-as-Judge quality scores (0–10) by category, showing score distributions across models. Black dots indicate means; horizontal lines indicate medians. \emph{Best view in Zoom.}
    }
    \label{fig:redsage-frq-eval-results}
\end{figure}

\textbf{Results on RedSage-Bench.} 
For MCQs, both base and instruction-tuned models are tested in the 0-shot setting, with Tab.~\ref{tab:redsage_mcq_results} showing that all RedSage variants outperform baselines across categories. For open-ended Q\&A, we evaluate instruction-tuned models using an LLM-as-Judge rubric to assess both factual correctness and answer quality (Sec.~\ref{subsubsec:redsagebench}). As shown in Fig.~\ref{fig:redsage-frq-eval-results}, RedSage achieves not only high accuracy but also the best answer quality across categories. 
More detailed results and qualitative examples illustrating model predictions are provided in Appendix \ref{app:redsage-benchmarks}.


\begin{figure*}[ht]
\centering \small
\begin{minipage}[t]{0.65\textwidth}
\centering
\vspace{-\baselineskip}
\captionof{table}{\small RedSage-MCQ (0-shot). Values are accuracy (\%). 
Abbreviations: Gen = General, Frm = Frameworks, 
Off = Offensive Skills, CLI = Command-line Tools, Kali = Kali Tools. 
Bold numbers indicate the best result \red{of 8B models}; underlined numbers indicate the second best.}
\label{tab:redsage_mcq_results}
\resizebox{\linewidth}{!}{%
\begin{tabular}{l c c c c c c}
\toprule
\rowcolor{Gray} 
\textbf{Model Name} & 
\multicolumn{1}{c}{\textbf{Macro}} & 
\multicolumn{2}{c}{\textbf{Knowledge}} & 
\multicolumn{1}{c}{\textbf{Skill}} & 
\multicolumn{2}{c}{\textbf{Tools}} \\
\cmidrule(lr){3-4} \cmidrule(lr){5-5} \cmidrule(lr){6-7}
 & \textbf{Acc} & \textbf{Gen} & \textbf{Frm} & \textbf{Off} & \textbf{CLI} & \textbf{Kali} \\
\midrule
\multicolumn{7}{c}{\textit{Base Model Evaluation (Text Completion)}} \\
\midrule
Llama-3.1-8B                    & 78.02 & 77.42 & 75.26 & 82.78 & 77.78 & 72.12 \\
Foundation-Sec-8B               & 78.51 & 76.82 & 79.10 & 83.68 & 76.64 & 71.14 \\
Qwen3-8B-Base                   & 84.24 & 83.08 & 81.94 & 88.23 & 85.08 & 78.86 \\
\midrule
\textbf{RedSage-8B-CFW}         & 84.86 & \underline{83.62} & 83.30 & \underline{88.81} & \underline{85.30} & 79.32 \\
\textbf{RedSage-8B-Seed}        & \textbf{85.21} & \textbf{83.64} & \underline{84.56} & \textbf{88.82} & \textbf{85.50} & \textbf{79.90} \\
\textbf{RedSage-8B-Base}        & \underline{85.05} & 83.12 & \textbf{84.94} & 88.72 & 85.44 & \underline{79.36} \\
\midrule
\multicolumn{7}{c}{\textit{Instruct Model Evaluation (w/ Chat Template)}} \\
\midrule
Lily-Cybersecurity-7B-v0.2      & 71.19 & 68.78 & 67.44 & 76.61 & 71.44 & 66.26 \\
Llama-Primus-Merged             & 74.81 & 74.34 & 72.34 & 79.31 & 74.74 & 68.82 \\
Foundation-Sec-8B-Instruct      & 76.12 & 74.50 & 77.10 & 80.91 & 74.98 & 68.30 \\
Llama-Primus-Base               & 77.02 & 76.78 & 74.10 & 80.87 & 76.78 & 72.72 \\
Llama-3.1-8B-Instruct           & 77.05 & 76.06 & 73.30 & 80.90 & 78.72 & 72.40 \\
DeepHat-V1-7B                   & 80.18 & 77.26 & 76.90 & 85.07 & 81.94 & 74.82 \\
Qwen3-8B                        & 81.85 & 80.46 & 78.82 & 86.16 & 83.92 & 75.56 \\
\midrule
\textbf{RedSage-8B-Ins}         &  \textbf{85.73} &  \textbf{84.20} &  \textbf{84.98} &  \textbf{89.06} &  \textbf{86.80} &  \textbf{80.30} \\
\textbf{RedSage-8B-DPO}         &  \underline{84.83} & \underline{82.48} & \underline{83.80} & \underline{88.54} & \underline{86.30} & \underline{79.30} \\
\midrule
\multicolumn{7}{c}{\textit{\red{Larger Instruct \& Proprietary Model Evaluation (w/ Chat Template)}}} \\
\midrule
Qwen3-32B                       & 85.40 & 84.08 & 82.32 & 89.00 & 87.60 & 80.40 \\
GPT-5                          & 88.68	& 88.74 & 86.54 & 91.43 & 90.80 & 83.14
 \\
\bottomrule
\end{tabular}
}
\end{minipage}\hfill
\begin{minipage}[t]{0.325\textwidth}
\small
\textbf{MCQ Analysis.}
Qwen3-8B-Base, trained on 36T tokens, is the strongest external \red{8B} baseline (84.24) and even outperforms Foundation-Sec-8B. underscoring the importance of selecting a strong base model. 
Building on it with CPT, RedSage gains up to +0.97 macro-accuracy points, with the largest improvements in \textit{Frameworks} (+3.00) and \textit{Kali} (+1.04). RedSage-8B-Seed achieves the best base result (85.21), demonstrating better alignment with the curated Seed data. 
Among instruction-tuned models, RedSage avoids the accuracy drop and exceeds Qwen3 by +2.98 (DPO) to +3.88 (Ins). DPO on \emph{general data} slightly lowers accuracy but stays well above baselines.
\red{Interestingly, RedSage-Ins surpasses Qwen3-32B on average despite its smaller size.}
These results highlight that our domain-aware CPT and SFT enhance robustness across cybersecurity knowledge, skills, and tools.

\end{minipage}
\end{figure*}

\textbf{Open-ended QA Analysis.}
RedSage-8B-DPO achieves the best performance (Fig.~\ref{fig:redsage-frq-eval-results}), surpassing the second-best model (Qwen3-8B) by +7\% absolute mean correctness and +0.07 in mean quality score. RedSage-8B-Ins attains similar correctness to Qwen3-8B but lags in answer quality (6.43), underscoring the role of preference alignment in producing not only accurate but also helpful responses. The remaining models fall substantially behind, with mean correctness ranging from 51\% to 40\% and quality scores from 5.84 to 4.28, highlighting a significant gap from the top three. The faceted violin plots further reveal category difficulty: knowledge tasks exhibit higher and tighter distributions, skill tasks lie in the middle range, and tool-use tasks show lower medians with heavy tails, pinpointing tool expertise as the primary challenge. These findings demonstrate the value of our benchmark for assessing cybersecurity capabilities in free-form answers.

\subsection{Evaluation Results on Cybersecurity Benchmarks}
\textbf{Results on Cybersecurity Benchmarks.} We assess generalization on multiple established benchmarks in Tab.~\ref{tab:related_cybersecurity_bench_combined}. 
For CyberMetric (CyMtc) \citep{cybermetric24}, we evaluate all models using the 500 human-verified MCQs. We select English (En) MCQs from SecBench (ScBen) \citep{jing2024secbench}. We also include MCQs related to the Computer Security (CSec) from MMLU \citep{hendryckstest2021}. 
For SECURE \citep{secure2024}, we evaluate models using the  MCQs types covering MAET, CWET, and KCV. Further, we evaluate all models on CTI-Bench \citep{ctibench2024} (MCQ, Root Cause Mapping (RCM)), and SecEval (ScEva) \citep{li2023seceval} (MCQ).
%
We provide further details about each benchmark and metrics in Appendix \ref{app:cybersecurity-benchmarks}.
Base models are evaluated with 5-shot prompting, and instruction-tuned models in 0-shot.

\begin{table*}[ht]
\centering \small
\caption{Benchmark results for Base and Instruct Models. 
Values are Accuracy (\%). Rows are sorted by mean performance. 
Best results \red{for 8B models} are in bold, second-best are underlined. }
\label{tab:related_cybersecurity_bench_combined}
\setlength{\tabcolsep}{5.5pt}
\scalebox{0.85}[0.85]{
\begin{tabular}{lcccccccccc}
\toprule
\rowcolor{Gray}
\multirow{2}{*}{\textbf{Model Name}} & \multirow{2}{*}{\textbf{Mean}} & \multicolumn{2}{c}{\textbf{CTI-Bench}} & \multicolumn{1}{c}{\textbf{CyMtc}} & \multicolumn{1}{c}{\textbf{MMLU}} & \multicolumn{1}{c}{\textbf{ScBen}} & \multicolumn{1}{c}{\textbf{ScEva}} & \multicolumn{3}{c}{\textbf{SECURE}} \\
\cmidrule(lr){3-4} \cmidrule(lr){5-5} \cmidrule(lr){6-6} \cmidrule(lr){7-7} \cmidrule(lr){8-8} \cmidrule(lr){9-11}
& & \textbf{MCQ} & \textbf{RCM} & \textbf{500} & \textbf{CSec} & \textbf{En} & \textbf{MCQ} & \textbf{CWET} & \textbf{KCV} & \textbf{MAET} \\
\midrule
\multicolumn{11}{c}{\textit{Base Model Evaluation (5-shot)}} \\
\midrule
Llama-3.1-8B              & 75.44 & 61.12 & 65.80 & 84.20 & 83.00 & 72.80 & 54.27 & 86.34 & 83.73 & 87.72 \\
Foundation-Sec-8B         & 76.90 & 62.40 & 75.40 & 86.60 & 80.00 & 69.86 & 55.64 & 88.01 & 84.38 & 89.78 \\
Qwen3-8B-Base             & 80.81 & 68.80 & 63.50 & 92.00 & 83.00 & 82.84 & 75.60 & 92.70 & 75.05 & 93.81 \\
\midrule
\textbf{RedSage-8B-CFW}   & 82.66 & 68.40 & 67.60 & \textbf{93.80} & 86.00 & \textbf{83.62} & \textbf{76.10} & \textbf{93.33} & 81.34 & 93.72 \\
\textbf{RedSage-8B-Seed}  & \underline{84.45} & \underline{70.80} & \textbf{78.60} & \underline{92.20} & \textbf{88.00} & 81.61 & \underline{75.96} & 93.12 & \underline{85.47} & \textbf{94.28} \\
\textbf{RedSage-8B-Base} & \textbf{84.56} & \textbf{71.04} & \underline{78.40} & 92.60 & \underline{87.00} & \underline{81.76} & 75.83 & \underline{93.22} & \textbf{87.20} & \underline{94.00} \\
\midrule
\multicolumn{11}{c}{\textit{Instruct Model Evaluation (0-shot)}} \\
\midrule
Lily-Cybersecurity-7B-v0.2             & 55.74 & 30.04 & 43.60 & 65.20 & 68.00 & 57.65 & 39.72 & 72.99 & 49.67 & 74.79 \\
Llama-3.1-8B-Instruct                  & 68.52 & 58.24 & 58.30 & 82.80 & 72.00 & 59.66 & 35.37 & 84.98 & 82.86 & 82.47 \\
Llama-Primus-Merged                    & 71.23 & 55.92 & 68.50 & 83.80 & 76.00 & 64.91 & 39.31 & 86.13 & 82.65 & 83.88 \\
Llama-Primus-Base                      & 71.69 & 52.32 & 68.50 & 83.80 & \textbf{79.00} & 63.68 & 61.15 & 88.01 & 65.08 & 83.69 \\
DeepHat-V1-7B                          & 75.44 & 62.08 & 68.20 & 86.00 & 74.00 & 70.63 & 56.65 & 87.07 & \underline{86.77} & 87.54 \\
Foundation-Sec-8B-Instruct             & 75.44 & 63.24 & 69.40 & 83.00 & 76.00 & 68.78 & 65.46 & 85.82 & 82.00 & 85.29 \\
Qwen3-8B                               & 75.71 & 62.76 & 54.00 & 88.60 & 76.00 & 73.26 & 65.46 & 88.11 & \textbf{87.42} & 85.75 \\
\midrule
\textbf{RedSage-8B-Ins}                & \textbf{81.30}    & \underline{70.56} & \textbf{76.70}    & \underline{89.80} & \underline{78.00} & \underline{79.91} & \underline{72.48} & \textbf{91.45} & 81.34 & \textbf{91.47} \\
\textbf{RedSage-8B-DPO}                & \underline{81.10} & \textbf{70.84}    & \underline{70.60} & \textbf{90.00}    & \textbf{79.00}    & \textbf{80.06}    & \textbf{74.22}    & \underline{91.35} & 82.86 & \underline{91.00} \\
\midrule
\multicolumn{11}{c}{\textit{\red{Larger Instruct and Proprietary Model Evaluation (0-shot)}}} \\
\midrule
Qwen3-32B                  & 82.31 & 70.04 & 65.60 & 91.80 & 84.00 & 84.23 & 76.23 & 89.46 & 89.37 & 90.06 \\
GPT-5                      & 86.29 & 76.48 & 74.20 & 95.60 & 86.00 & 87.48 & 83.03 & 92.70 & 88.72 & 92.41 \\

\bottomrule
\end{tabular}}
\end{table*}


\textbf{Analysis.}
Across related cybersecurity benchmarks, RedSage base models improve over Qwen3-8B-Base (80.81\%) by up to +3.75 points. CPT with CFW leads on SecBench (83.62), CyMtc (93.80), and CWET (93.33), raising the mean by +1.85. CPT with Seed excels on CTI-RCM (78.60), MMLU-CSec (88.00), and MAET (94.28), lifting the mean by +3.64. Combining both yields the best overall mean (84.56) and top scores on CTI-MCQ (71.04) and KCV (87.20). In the 0-shot instruct setting, RedSage surpasses Qwen3-8B (75.71\%) by +5.39 (DPO) to +5.59 (Ins). Except for Lily-Cybersecurity, all domain-tuned baselines outperform Llama-3.1-8B-Instruct, though still lag behind RedSage. 
\red{Despite having far fewer parameters, RedSage comes close to Qwen3-32B (82.31 mean, only about +1 point higher) and trails GPT-5 (86.29 mean, roughly +5 points higher), highlighting strong efficiency relative to much larger models.}
These results show that CyberFineWeb and Seed provide complementary strengths, while different post-training strategies specialize across tasks, together setting new state-of-the-art performance in cybersecurity LLM evaluation.

\subsection{Evaluation Results on General Benchmarks}

We use benchmarks from the Open LLM Leaderboard in Lighteval, including ARC-Challenge (ARC-C)~\citep{Clark2018ThinkYH}, HellaSwag (HSwag)~\citep{zellers2019hellaswag}, TruthfulQA (TQA)~\citep{lin2021truthfulqa}, MMLU~\citep{hendryckstest2021}, WinoGrande (WinoG)~\citep{sakaguchi2020winogrande}, GSM8K~\citep{cobbe2021gsm8k}, and IFEval~\citep{zhou2023ifeval}.
Results in Tab.~\ref{tab:open_llm_leaderboard_combined} show our instruction-tuned models achieve competitive results on general tasks, surpassing baselines by a clear margin. Benchmark configurations and evaluation metrics are provided in Appendix \ref{app:general-benchmarks}.

\begin{table*}[ht]
\centering
\caption{Open LLM Leaderboard Benchmarks. All values are accuracy (\%). Bold numbers indicate the best result \red{for 8B models} and underlined numbers indicate the second best.}
\label{tab:open_llm_leaderboard_combined}
\setlength{\tabcolsep}{8pt}
		\scalebox{0.80}[0.80]{
\begin{tabular}{lcccccccc}
\toprule
\rowcolor{Gray} 
\textbf{Model Name} & \textbf{Mean} & \textbf{MMLU} & \textbf{ARC-C} & \textbf{GSM8K} & \textbf{HSwag} & \textbf{TQA} & \textbf{WinoG} & \textbf{IFEvl} \\
\midrule
\multicolumn{9}{c}{\textit{Base Model Evaluation (Mean excludes IFEval)}} \\
\midrule
Llama-3.1-8B                   & 61.15 & 66.31 & 58.19 & 49.05 & \textbf{82.08} & 35.98 & \textbf{75.30} & \textemdash \\
Foundation-Sec-8B              & 60.24 & 63.62 & 58.45 & 46.17 & 81.32 & 38.71 & \underline{73.16} & \textemdash \\
Qwen3-8B-Base                  & \textbf{70.86} & \textbf{78.73} & \textbf{68.09} & 81.73 & \underline{79.62} & \textbf{43.84} & \underline{73.16} & \textemdash \\
\midrule
\textbf{RedSage-8B-CFW}        & 69.31 & \underline{78.63} & \underline{66.72} & 81.12 & 79.26 & 38.09 & 72.06 & \textemdash \\
\textbf{RedSage-8B-Seed}       & \underline{69.58} & 78.18 & 65.19 & \textbf{82.34} & 77.76 & \underline{42.44} & 71.59 & \textemdash \\
\textbf{RedSage-8B-Base}       & 69.23 & 77.80 & 65.53 & \underline{82.03} & 77.96 & 42.19 & 69.85 & \textemdash \\
\midrule
\multicolumn{9}{c}{\textit{Instruct Model Evaluation (Mean includes IFEval)}} \\
\midrule
Lily-Cybersecurity-7B-v0.2     & 56.98 & 56.49 & 58.96 & 30.86 & 80.94 & 48.53 & 72.06 & 50.99 \\
Llama-Primus-Base              & 64.82 & 65.09 & 51.19 & 71.80 & 79.49 & 44.62 & 72.69 & 68.85 \\
DeepHat-V1-7B                  & 64.89 & 69.53 & 57.17 & 77.94 & 74.80 & 33.17 & 69.06 & 72.58 \\
Qwen3-8B                       & 65.92 & 73.59 & 62.54 & 75.66 & 56.70 & 45.23 & 62.51 & \textbf{85.21} \\
Llama-Primus-Merged            & 66.71 & 66.17 & 53.07 & 75.28 & 79.07 & 46.52 & \underline{73.24} & 73.58 \\
Llama-3.1-8B-Instruct          & 68.20 & 67.29 & 57.51 & 77.41 & 78.91 & 45.93 & 72.61 & 77.75 \\
Foundation-Sec-8B-Instruct     & 69.28 & 64.11 & 63.91 & 77.79 & \textbf{81.35} & \textbf{53.15} & 68.51 & 76.17 \\
\midrule
\textbf{RedSage-8B-Ins}        & \underline{73.34} & \textbf{77.38} & \underline{69.62} & \textbf{86.05} & 79.00 & 47.75 & \textbf{73.64} & 79.97 \\
\textbf{RedSage-8B-DPO}        & \textbf{74.33} & \underline{77.07} & \textbf{71.76} & \underline{82.71} & \underline{79.87} & \underline{52.47} & 73.01 & \underline{83.44} \\
\midrule
\multicolumn{9}{c}{\textit{\red{Larger Instruct and Proprietary Model Evaluation}}} \\
\midrule
Qwen3-32B	         & 73.17 & 82.11 &  69.28 & 87.49 & 70.93 & 48.17 & 65.98 & 88.26 \\
GPT-5                & 91.07 & 91.4 & 95.31 & 91.36 & 94.85 & 87.10 & 87.85 & 89.60 \\
\bottomrule
\end{tabular}}
\caption*{\footnotesize IFEval (Instruction-Following Eval) is excluded from base models as it is designed for instruct-tuned models.}
\vspace{-1.5em}
\end{table*}

\textbf{Analysis.}
Among base models, Qwen3-8B-Base is strongest overall (70.86) and leads MMLU (78.73) and ARC-C (68.09), while Llama-3.1-8B tops HSwag (82.08) and WinoG (75.30). RedSage bases are competitive in mean (69.23–69.58) and achieve task highs, including best GSM8K (82.34, Seed) and second on MMLU (78.63, CFW) and ARC-C (66.72, CFW), where the slight drop may stem from our FineWeb-Edu general-knowledge replay strategy. After instruction tuning, RedSage attains the best overall mean with DPO (74.33) and second with Ins (73.34), setting new highs on ARC-C (71.76, DPO), GSM8K (86.05, Ins), MMLU (77.38, Ins), and leading WinoG (73.64, Ins). 
Foundation-Sec-8B-Instruct leads HSwag (81.35) and TQA (53.15), and Qwen3-8B leads IFEval (85.21), with RedSage-DPO close (83.44). 
\red{For larger and proprietary models, the performance gap widens: GPT-5 reaches a 91.07 mean accuracy, but RedSage-8B-DPO still surpasses Qwen3-32B (74.33 vs.\ 73.17) due to gains on HellaSwag, TQA, and WinoGrande, which emphasize commonsense reasoning and factuality.}
These patterns indicate complementary effects: Seed boosts math reasoning (GSM8K), CFW strengthens general knowledge and reasoning
(MMLU and ARC-C), and DPO improves instruction-following 
(IFEvl), while RedSage remains competitive on general tasks despite cybersecurity tuning. \red{Importantly, the 8B-scale RedSage model can be deployed locally on consumer-grade GPUs, enabling privacy-preserving on-premise use.}

\section{Discussion and Limitations}



The data pipeline, which leverages LLM-generated content and verification, scales effectively but may still propagate biases or inaccuracies despite screening. Nevertheless, our benchmark extends existing cybersecurity evaluations, fills missing dimensions, and offers value to the community.
Finally, as the model incorporates offensive security knowledge, it carries an inherent risk of misuse. While such dual-use concerns are intrinsic in cybersecurity research, we emphasize the importance of responsible application and good security practices to promote ethical use.

\section{Conclusion}

We presented \textsc{RedSage}, an open cybersecurity assistant that combines a large-scale pretraining corpus (\textsc{CyberFineWeb}, 11.7B tokens), a curated seed of authoritative resources (\textsc{RedSage-Seed}, 29K items, 150M tokens), and 266K augmented dialogues for supervised fine-tuning, together with \textsc{RedSage-Bench}, a 30K-question benchmark spanning knowledge, skills, and tool use. At the 8B scale, \textsc{RedSage} achieves state-of-the-art results, surpassing baselines by up to +5.9 points on cybersecurity tasks and +5.0 on general LLM benchmarks, while avoiding the post-tuning degradation observed in prior models. \red{Because RedSage runs at 8B, it supports privacy-preserving, on-prem deployment on consumer-grade GPUs, enabling practical use without relying on cloud inference.}
We will release all models, datasets, and code to support reproducibility and accelerate open research on practical and domain-specialized AI assistants for cybersecurity.

\section{Ethics Statement}

This work adheres to the ICLR Code of Ethics. All datasets used in this study were derived exclusively from publicly available and internet-accessible sources. Our large-scale pretraining corpus builds directly on prior work that already applied extensive filtering, deduplication, and removal of personally identifiable information (PII). We further applied additional quality checks to ensure that the data contain only non-sensitive and appropriately licensed content.

We note that some components of the curated \textsc{RedSage} datasets may include publicly available but copyrighted resources (e.g., educational portals, online tutorials, or news articles). Such content was used solely for non-commercial academic research, and we will not redistribute these resources without obtaining the necessary permissions from the rights holders. Only aggregated statistics are reported in this paper, and any public release of datasets will exclude copyrighted material unless explicit approval has been secured.

As part of the writing process, we used large language models responsibly and only for editorial assistance (e.g., polishing phrasing, improving readability, and checking grammar).

The \textsc{RedSage} models are released strictly for research purposes and not intended for deployment in real-world security operations without additional safeguards. To support responsible use, we will make models, datasets, and code openly available under research-friendly licenses with clear documentation and usage guidelines, promoting transparency, reproducibility, and community benefit.

\section{Reproducibility Statement}

We are committed to advancing reproducibility and open research in cybersecurity-oriented LLMs by releasing our datasets, models, and code. The collection and augmentation of our datasets for domain-aware pre- and post-training are described in Sec.~\ref{section:proposal}, with detailed descriptions, statistics, and implementation details (including prompt templates) provided in Appendix~\ref{app:dataset-details}. Model training procedures are presented in Sec.~\ref{subsection:model_training}, with further implementation details in Appendix~\ref{app:training-details}. 

Our models are trained using the Axolotl framework~\citep{axolotl}, which facilitates direct replication through reusable configuration files; users need only replace the base model and dataset. All hyperparameters are fully specified in Appendix~\ref{app:training-details}. For evaluation, we use the Hugging Face LightEval framework~\citep{lighteval} to run all benchmarks, ensuring reproducibility and enabling evaluation of arbitrary LLMs by specifying the benchmark configuration. Our evaluation protocol, compared models, and benchmark details are documented in Sec.~\ref{section:experiments} and Appendix~\ref{app:evaluation-details}. All datasets, code, and evaluation pipelines are released as open-source.

\section*{Acknowledgments}
This research was funded by Khalifa University of Science and Technology through the Faculty Start-Ups under Project ID: KU-INT-FSU-2005-8474000775. Further, the work has been supported by the Federal Ministry of Research, Technology and Space (BMFTR) under grant no. 16IS22094A WEST-AI. For the computations involved in this research, we acknowledge EuroHPC Joint Undertaking for awarding us access to Leonardo at CINECA, Italy, through EuroHPC Regular Access Call proposal No. EHPC-REG-2024R02-192.

\bibliography{main_preprint}
\bibliographystyle{main_preprint}

\newpage

\appendix

\section{Dataset Details}
\label{app:dataset-details}

This section details the datasets we created and curated for training our LLM. All token counts are computed with the GPT-2 tokenizer\footnote{GPT-2: \href{https://huggingface.co/openai-community/gpt2}{openai-community/gpt2}}, following the conventions of FineWeb \citep{penedo2024the}.

\subsection{CyberFineWeb}
\label{app:cyberfineweb}

\textbf{CyberFineWeb} is derived from the original FineWeb dataset \citep{penedo2024the}\footnote{FineWeb Datasets: \href{https://huggingface.co/datasets/HuggingFaceFW/fineweb}{HuggingFaceFW/fineweb}}, a large-scale, cleaned web corpus aggregated from Common Crawl. Although FineWeb is continuously updated, for our development we used all subsets released between Summer 2013 (CC-MAIN-2013-20) and December 2024 (CC-MAIN-2024-51). This selection comprises 104 subsets, totaling 46,934 GB of data and 17.2 trillion tokens.

\textbf{Text Classification Model} 
To extract the cybersecurity corpus from FineWeb, we trained a text classification model based on ModernBERT-base~\citep{warner2024smarter}, a state-of-the-art transformer encoder. The training data came from the Cybersecurity Topic Classification dataset \citep{pelofske2021robust}, which contains 9.27M labeled training samples (cybersecurity vs. non-cybersecurity) collected from Reddit, StackExchange, and arXiv, along with 459K validation samples from web articles. 
\red{The labels in this dataset originate from forum categories, tags, and keyword metadata rather than from LLM-generated annotations.}
To reduce context ambiguity, we filtered out very short texts, yielding 4.62M training samples and 2.46K validation samples. 
The model was trained with the Adam optimizer for 2 epochs using a learning rate of 2e-5 and a 10\% warmup ratio. 
On the validation set, the model achieved 93.8\% precision, 90.2\% recall,  91.4 \% F1 score and 97.3\% accuracy.

\begin{figure}[h]
    \centering
    \includegraphics[width=\linewidth]{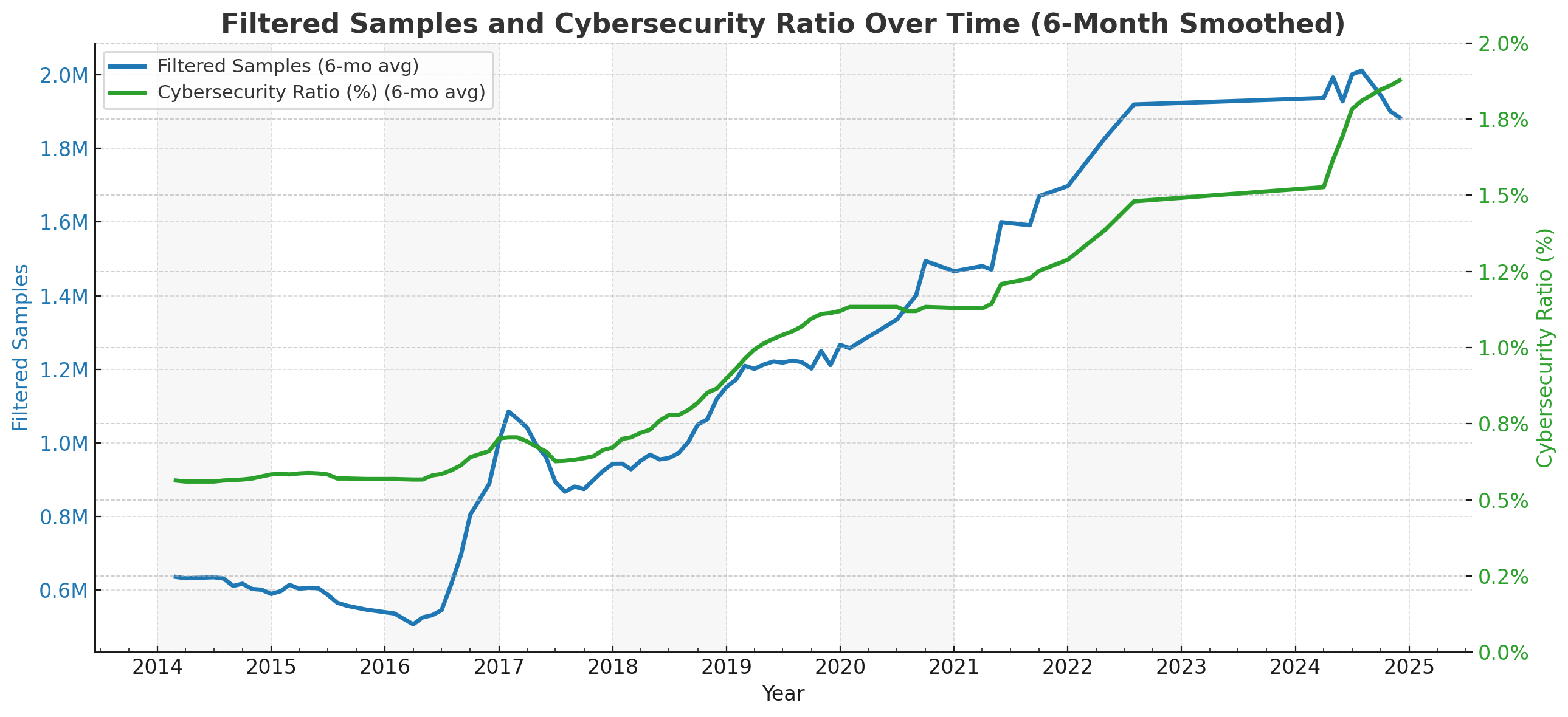}
    \caption{Number of filtered cybersecurity samples and their ratio over time across FineWeb subsets.}
    \label{fig:fineweb-filtering-ratio}
\end{figure}

\textbf{Text Filtering} 
We applied the trained classifier to each subset of FineWeb. 
Figure \ref{fig:fineweb-filtering-ratio} shows the number of identified cybersecurity samples and their relative proportion across all subsets ordered by crawl date. The results indicate a steady increase in cybersecurity-related content on the web, underscoring the growing importance of this domain. 
In total, this filtering process produced approximately 125M documents ($\sim$89.8B tokens), corresponding to about 0.77\% of the original FineWeb.

\textbf{General Knowledge Integration} 
Due to compute constraints, we partitioned the dataset into 20 chronological chunks. To mitigate catastrophic forgetting of general-domain knowledge, we first select a fixed 100B-token subset from FineWeb-Edu \citep{lozhkov2024fineweb-edu}. For each chunk, we then randomly resampled data from this subset to match 30\% of the chunk's size, ensuring balanced exposure to general-domain content throughout training.

\textbf{Deduplication} 
Although FineWeb includes text deduplication in its pipeline, it is applied only within individual CommonCrawl dumps. We applied global deduplication across our mixed corpus using MinHash-LSH implemented in DataTrove \citep{penedo2024datatrove}, with 64-bit precision, 14 buckets, and 8 hashes per bucket. This reduced the corpus size by 58.4\% in documents (to $\sim$52M) and by 47.9\% in tokens (to $\sim$46.8B).

\textbf{Final Corpus} 
To fit our training budget, we selected the latest 5 chunks from the mixed, deduplicated data. This formed our final pretraining corpus, containing $\sim$13M documents ($\sim$11.7B tokens). A summary of the dataset filtering and processing steps from FineWeb to the final CyberFineWeb corpus is provided in Table~\ref{tab:dataset-summary}.

\begin{table}[h]
\centering
\caption{Summary of dataset filtering and processing stages from FineWeb to the final CyberFineWeb corpus. Retention percentages are relative to the original FineWeb.}
\label{tab:dataset-summary}
\setlength{\tabcolsep}{7pt}
		\scalebox{0.90}[0.90]{
\begin{tabular}{lrrr}
\toprule
\rowcolor{Gray} 
\textbf{Stage} & \textbf{Documents} & \textbf{Tokens} & \textbf{Retention (vs. FineWeb)} \\
\midrule
FineWeb (2013-2024, 104 subsets) & $\sim$24.5B & $\sim$17.2T & 100\% \\
CyberFineWeb (after filtering) & $\sim$125M & $\sim$89.8B & 0.51\% docs / 0.52\% tokens \\
General-mixing + deduplication (20 chunks) & $\sim$52M & $\sim$46.8B & 0.21\% docs / 0.27\% tokens \\
Final CyberFineWeb corpus (latest 5 chunks) & $\sim$13M & $\sim$11.7B & 0.053\% docs / 0.068\% tokens \\
\bottomrule
\end{tabular}%
}
\end{table}

\subsection{RedSage Seed}
\label{app:redsageseed}

\textbf{RedSage Seed.} Our curated collection of publicly available cybersecurity resources is designed to provide high-quality pretraining data in structured Markdown format. We excluded private resources such as books to ensure that all data are openly accessible. 

Some resources, such as MITRE ATT\&CK, CAPEC, and CWE \citep{mitre_attack, mitre_capec, mitre_cwe}, are distributed as XML files, which we parsed into structured Markdown while preserving the original website organization. Other resources, such as \texttt{tldr-pages} \citep{tldr_pages} and \texttt{kali-tools} \citep{kali_tools}, were already available in Markdown format. For curated webpages, we crawled and processed them using Jina ReaderLM-v2~\citep{wang2025readerlmv2} to convert the HTML content into Markdown. 

The RedSage-Seed corpus is organized into three main categories: \emph{knowledge}, \emph{skills}, and \emph{tools}. 
Within \textbf{knowledge}, we distinguish between (i) \emph{General}, which includes sources such as Wikipedia and roadmap.sh \citep{roadmapsh_cybersecurity}, and (ii) \emph{Frameworks}, which cover structured knowledge bases from MITRE and the OWASP Foundation \citep{owasp_top10}. 
For \textbf{skills}, we currently focus on offensive security, curating resources such as offensive tricks \citep{hacktricks}, articles \citep{hackingarticles}, community tutorial \citep{nullbyte}, and CTF write-ups \citep{0df}. 
Finally, \textbf{tools} are divided into (i) \emph{CLI}, which includes multi-platform command-line resources such as \texttt{tldr-pages} \citep{tldr_pages} and Unix man pages, and (ii) \emph{Kali Linux Tools} \citep{kali_tools}, which provide documentation for a curated set of cybersecurity tools. 
Dataset statistics and detailed categorization are presented in Table~\ref{tab:redsage_seed_stats_by_cat}. These resources also serve as the foundation for our agentic augmented cybersecurity conversations and benchmarking.

\textbf{RedSage Dump.} 
To complement RedSage-Seed and expand the diversity of high-quality data for cybersecurity pretraining, we curated additional publicly available resources under the RedSage Dump collection. 
This corpus aggregates technical documents, standards, and domain-specific reports that are particularly relevant for developing a cybersecurity assistant. 
Specifically, it includes: (i) \emph{Computer Education Portals} \citep{geeksforgeeks}, which provide structured tutorials and training materials on computer science and cybersecurity fundamentals; (ii) \emph{Cybersecurity News} \citep{cybersecurity_news}, capturing timely reports and analyses of emerging threats and incidents; (iii) \emph{RFC Entries} \citep{rfc_entries}, representing standardized internet protocols and technical specifications; (iv) \emph{NIST Publications} \citep{nist_publications}, offering authoritative cybersecurity and compliance guidelines; (v) \emph{Primus Seed} \citep{yu2025primus}, a curated collection of cybersecurity resources originally used to pretrain the Primus model; and (vi) the \emph{National Vulnerability Database (NVD)} \citep{nvd}, which provides structured vulnerability advisories. 

Statistics for these sources are summarized in Table~\ref{tab:redsage_dump}. 
Overall, the RedSage Dump contains 459K documents with a total of $\sim$700M tokens. 
This collection complements RedSage-Seed by emphasizing technical standards, structured vulnerability data, and up-to-date cybersecurity reporting.

\begin{table*}[h]
\centering
\caption{RedSage Seed Statistics by Category: Samples and Tokens}
\label{tab:redsage_seed_stats_by_cat}
\setlength{\tabcolsep}{9pt}
		\scalebox{0.850}[0.850]{
\begin{tabular}{lrrrrr}
\toprule
\rowcolor{Gray} 
\textbf{Configuration} & \textbf{Samples} & \textbf{Avg. Tokens} & \textbf{Total Tokens} & \textbf{Min Tokens} & \textbf{Max Tokens} \\
\midrule
\multicolumn{6}{l}{\textbf{Knowledge -- General}} \\
Cybersecurity Wikis     & 6,636 & 2,304.77 & 15,294,454 & 39  & 36,812  \\
Cybersecurity Roadmaps  &   288 & 3,671.35 &  1,057,349 & 86  & 171,839 \\
\midrule
\multicolumn{6}{l}{\textbf{Knowledge -- Frameworks}} \\
MITRE ATT\&CK       & 1,655 & 4,806.38 & 7,954,559 & 366 & 96,808 \\
MITRE CAPEC         &   589 &   654.42 &   385,453 & 61  & 2,444 \\
MITRE CWE           & 1,346 & 1,222.46 & 1,645,431 & 140 & 10,679 \\
OWASP               &   125 & 4,313.63 &   539,204 & 436 & 17,253 \\
\midrule
\multicolumn{6}{l}{\textbf{Skill -- Offensive}} \\
Offensive Tricks        & 1,050 & 2,924.06 & 3,070,263 & 116 & 29,902 \\
Hacking Articles        & 1,384 & 13,919.66 & 19,264,809 & 377 & 190,391 \\
Null Byte Tutorials     & 1,002 & 4,402.07 & 4,410,874 & 278 & 79,225 \\
CTF Write-ups           &   596 & 18,471.77 & 11,009,175 & 185 & 83,759 \\
\midrule
\multicolumn{6}{l}{\textbf{Tools -- CLI}} \\
TLDR Pages (English)    & 5,335 & 11,215.81 & 59,836,346 & 35  & 543,349 \\
Unix Man Pages          & 7,608 & 2,509.00  & 19,088,472 & 45  & 379,876 \\
\midrule
\multicolumn{6}{l}{\textbf{Tools -- Kali}} \\
Kali Documentation      &   265 & 1,568.08 &   415,541 & 53  & 17,983 \\
Kali Tools              &   758 & 7,722.30 & 5,853,503 & 169 & 709,750 \\
\midrule
\textbf{Total (dataset)} & \textbf{28,637} & \textbf{5,231.00} & \textbf{149,825,433} & \textbf{35} & \textbf{709,750} \\
\bottomrule
\end{tabular}}
\end{table*}

\begin{table*}[ht]
\centering
\caption{RedSage Dump Statistics}
\label{tab:redsage_dump}
\setlength{\tabcolsep}{15pt}
		\scalebox{0.90}[0.90]{
\begin{tabular}{lrrr}
\toprule
\rowcolor{Gray} 
\textbf{Source} & \textbf{Samples} & \textbf{Avg. Tokens} & \textbf{Total Tokens} \\
\midrule
Computer Education Portals     & 160,355 & 1,986 & 318,503,184 \\
Cybersecurity News      &  13,959 & 1,431 &  19,968,138 \\
RFC Entries             &   9,674 & 20,994 & 203,093,862 \\
NIST Publications       &   1,015 & 29,715 &  30,161,170 \\
Primus Seed (Website, Mitre)  &  80,336 &   849 &  68,233,498 \\
National Vulnerability Database (NVD) & 194,134 &   310 &  60,173,508 \\
\midrule
\textbf{Total}          & \textbf{459,473} & \textbf{1,524} & \textbf{700,133,360} \\
\bottomrule
\end{tabular}}
\end{table*}



\subsection{RedSage Conversation}
\label{app:redsageconv}

\textbf{Agentic Data Augmentation.} 
Our supervised finetuning (SFT) cybersecurity datasets are generated using an agentic augmentation pipeline. 
We first segment the RedSage-Seed corpus into chunks of up to 32,768 tokens using a Markdown text splitter. 
These chunks serve as the input to the planner agent, which determines appropriate augmentation strategies. \red{Within this pipeline, we adopt Llama-3.3-70B as the teacher model, as it was among the strongest open-source instruction-tuned models that could be run locally given our available compute during the data creation phase.}

\textbf{Planner Agent.} 
For each seed data chunk, the planner agent analyzes the content and proposes multiple skill sets, each associated with one or more augmentation types and descriptive transformations. 
This design enables diverse augmentation paths from the same source material, ensuring broad coverage of cybersecurity skills and tasks. 
Below is our planner agent's system prompt.

\begin{PromptBox}[Planner Agent's System Prompt]
\small
You are an \textbf{Augmentation Type Planner Agent} specializing in cybersecurity and penetration testing. 
Your role is to analyze a provided chunk of seed data and produce a structured, comprehensive list of possible skill sets and augmentation types. 
The resulting suggestions will be used by a \textbf{Data Augmentation Agent} to generate conversational training data for a chatbot. 
Keep in mind that the final output should lend itself well to turn-based dialogues, persona-based Q\&A, or scenario simulations typical in a chatbot environment.

\textbf{Objective}: \\
Given a chunk of \textbf{preprocessed seed data} related to cybersecurity and penetration testing, generate a well-structured list of \textbf{skill sets} and corresponding \textbf{augmentation types}. 
The suggestions should improve dataset quality, diversity, and relevance, and be easily adaptable into a conversational format (e.g., question--answer pairs, scenario-based dialogues, guided reasoning steps). 
Leverage the seed data's domain context to ensure accuracy and practical utility.

\textbf{Input}: 
\begin{itemize}[leftmargin=1.25em]
  \item \textbf{Seed Data}: A chunk of preprocessed markdown-formatted data related to cybersecurity and penetration testing.
\end{itemize}

\textbf{Output}: 
\begin{itemize}[leftmargin=1.25em]
  \item \textbf{Structured List of Skill Sets and Augmentation Types (in JSON format)}:
  \begin{itemize}
    \item Include multiple skill sets, each with several augmentation types.
    \item For each augmentation type, provide a brief description that clarifies its intended transformation and explains how it could be adapted into a conversational format for a chatbot.
  \end{itemize}
\end{itemize}

\textbf{Guidelines}:
\begin{enumerate}[leftmargin=1.25em]
  \item \textbf{Relevance and Grounding}: 
    \begin{itemize}
      \item Ensure all skill sets and augmentation types are relevant to cybersecurity and penetration testing.
      \item Ground the augmentations in the seed data to maintain accuracy. If the seed data mentions specific tools, vulnerabilities, or scenarios, align the augmentation suggestions accordingly.
    \end{itemize}
  \item \textbf{Diversity and Coverage}: 
    \begin{itemize}
      \item Suggest a wide range of augmentation strategies reflecting various penetration testing phases: reconnaissance, enumeration, exploitation, post-exploitation, mitigation, and so forth.
      \item Include traditional data transformations (e.g., paraphrasing) and advanced, scenario-based augmentations (e.g., simulating a penetration test conversation between a tester and a security analyst).
    \end{itemize}
  \item \textbf{Conversational Adaptability}: 
    \begin{itemize}
      \item Consider how each augmentation could be represented in a chatbot-friendly format (e.g., multi-turn Q\&A, narrative scenarios, role-based conversations, or step-by-step reasoning).
      \item Example: a vulnerability analysis could become a Q\&A where the chatbot explains the vulnerability to a novice, or a roleplay between red-teamer and blue-teamer discussing mitigation.
    \end{itemize}
  \item \textbf{Creativity and Innovation}: 
    \begin{itemize}
      \item Introduce new skill sets or augmentation ideas beyond predefined examples.
      \item Encourage creative transformations that leverage the chatbot setting (persona-based coaching, guided threat mapping dialogues, multi-turn explorations).
    \end{itemize}
  \item \textbf{Detail and Clarity}: 
    \begin{itemize}
      \item Each augmentation type should have a short description explaining what it does, how it relates to the seed data, and how it can be adapted into a conversational format.
    \end{itemize}
  \item \textbf{Format Requirements}: 
    \begin{itemize}
      \item Return output in JSON format with the following structure:
\begin{lstlisting}[language=JSON,showstringspaces=false]
{
  "skill_sets": [
    {
      "name": "<Skill Set Name>",
      "augmentation_types": [
        {
          "type": "<Descriptive Augmentation Type Name or Title>",
          "description": "<Brief explanation>"
        },
        {
          "type": "<Another Augmentation Type>",
          "description": "<Explanation>"
        }
      ]
    }
  ]
}
\end{lstlisting}
    \end{itemize}
\end{enumerate}

\textbf{Example Categories for Inspiration (Not Exhaustive)}:
\begin{itemize}[leftmargin=1.25em]
  \item \textbf{Tool Command Generation}: Turn tool usage into Q\&A or guided instructions.
  \item \textbf{Vulnerability Analysis}: Conversational explanations (senior engineer teaching junior tester).
  \item \textbf{Attack Surface Enumeration}: Brainstorming dialogue between red teamers.
  \item \textbf{Exploitation Workflows}: Multi-turn walkthroughs of exploitation stages.
  \item \textbf{Post-Exploitation Techniques}: Conversations explaining persistence and mitigation.
  \item \textbf{Threat Mapping}: Align findings with MITRE ATT\&CK in dialogue form.
  \item \textbf{Mitigation Recommendations}: Remediation advice in Q\&A format.
  \item \textbf{Scenario Simulation}: Role-based pen-test scenarios (e.g., tester vs client rep).
  \item \textbf{Step-by-Step Reasoning}: Guided reasoning revealed turn by turn.
  \item \textbf{Edge-Case Generation}: Ambiguity/error handling dialogues.
  \item \textbf{Reading Comprehension}: Extracting and explaining vulnerability advisories.
\end{itemize}

\textbf{Task}: \\
Use the provided seed data to produce a JSON-structured list of skill sets and augmentation types. 
Ensure augmentations can be easily converted into conversational formats. 
Suggest as many diverse and creative transformations as possible. 
The output must follow the specified JSON format; otherwise, it will be rejected.
\end{PromptBox}

For example, given the following seed data:

\begin{PromptBox}[arp-scan]
\small
\textbf{Homepage:} \url{https://github.com/royhills/arp-scan}\\
\textbf{Repository:} \url{https://salsa.debian.org/pkg-security-team/arp-scan}\\
\textbf{Architectures:} any\\
\textbf{Version:} 1.10.0-2\\
\textbf{Metapackages:} kali-linux-default, kali-linux-everything, kali-linux-headless, kali-linux-large

\medskip
\textbf{arp-scan}\\
arp-scan is a command-line tool that uses the ARP protocol to discover and
fingerprint IP hosts on the local network. It is available for Linux and BSD
under the GPL licence.

\textbf{Installed size:} \texttt{1.53 MB}\\
\textbf{How to install:} \texttt{sudo apt install arp-scan}

\medskip
\textbf{Dependencies:}
\begin{itemize}[leftmargin=1.25em]
  \item libc6
  \item libcap2
  \item libpcap0.8t64
\end{itemize}

\medskip\hrule\medskip

\textbf{arp-fingerprint}\\
Fingerprint a system using ARP

\begin{lstlisting}
root@kali:~# arp-fingerprint -h
Usage: arp-fingerprint [options] <target>
Fingerprint the target system using arp-scan.

'options' is one or more of:
        -h Display this usage message.
        -v Give verbose progress messages.
	-o <option-string> Pass specified options to arp-scan
	-l Fingerprint all targets in the local net.
\end{lstlisting}

\medskip\hrule\medskip

\textbf{arp-scan}\\
Send ARP requests to target hosts and display responses

\begin{lstlisting}[
  breaklines=true,
  breakatwhitespace=true,
  showstringspaces=false, 
]
root@kali:~# arp-scan -h
Usage: arp-scan [options] [hosts...]

Target hosts must be specified on the command line unless the --file or
--localnet option is used.

arp-scan uses raw sockets, which requires privileges on some systems:

Linux with POSIX.1e capabilities support using libcap:
       arp-scan is capabilities aware. It requires CAP_NET_RAW in the permitted
       set and only enables that capability for the required functions.
BSD and macOS:
       You need read/write access to /dev/bpf*
Any operating system:
       Running as root or SUID root will work on any OS but other methods
       are preferable where possible.

Targets can be IPv4 addresses or hostnames. You can also use CIDR notation
(10.0.0.0/24) (network and broadcast included), ranges (10.0.0.1-10.0.0.10),
and network:mask (10.0.0.0:255.255.255.0).

Options:

The data type for option arguments is shown by a letter in angle brackets: 

<s> Character string.
<i> Decimal integer, or hex if preceeded by 0x e.g. 2048 or 0x800.
<f> Floating point decimal number.
<m> MAC address, e.g. 01:23:45:67:89:ab or 01-23-45-67-89-ab (case insensitive)
<a> IPv4 address e.g. 10.0.0.1
<h> Hex encoded binary data. No leading 0x. (case insensitive).
<x> Something else - see option description.

General Options:

--help or -h		Display this usage message and exit.

--verbose or -v		Display verbose progress messages.
			Can be used than once to increase verbosity. Max=3.

--version or -V		Display program version details and exit.
			Shows the version, license details, libpcap version,
			and whether POSIX.1e capability support is included.

--interface=<s> or -I <s> Use network interface <s>.
			If this option is not specified, arp-scan will search
			the system interface list for the lowest numbered,
			configured up interface (excluding loopback).

Host Selection:

--file=<s> or -f <s>	Read hostnames or addresses from the specified file
			One name or address pattern per line. Use "-" for stdin.

--localnet or -l	Generate addresses from interface configuration.
			Generates list from interface address and netmask
			(network and broadcast included). You cannot use the
			--file option or give targets on the command line.
			Use --interface to specify the interface.
            
[... Data is trimmed for visualization ...]
\end{lstlisting}

\end{PromptBox}

The Planer Agent will output the following JSON:

\begin{PromptBox}[Beautified JSON Example]
\begin{lstlisting}[
  language=JSON,
  breaklines=true,
  breakatwhitespace=true,
  columns=fullflexible,   
  showstringspaces=false, 
basicstyle=\ttfamily\footnotesize
]
{
  "skill_sets": [
    {
      "name": "Network Discovery",
      "augmentation_types": [
        {
          "type": "ARP Scan Simulation",
          "description": "Simulate an ARP scan on a local network, displaying IP and MAC addresses of responding hosts, adaptable into a Q&A format where the chatbot explains the process and results."
        },
        {
          "type": "Host Fingerprinting",
          "description": "Transform host fingerprinting data into a conversational explanation, where a chatbot character educates another on operating system and device identification techniques."
        }
      ]
    },
    {
      "name": "Tool Command Generation",
      "augmentation_types": [
        {
          "type": "arp-scan Command Crafting",
          "description": "Generate `arp-scan` commands based on user input (e.g., IP range, interface), and have the chatbot walk the user through command construction and execution."
        },
        {
          "type": "Custom Vendor Mapping",
          "description": "Create a conversation where the chatbot helps a user understand and customize vendor mapping files for `arp-scan`, including explanations of MAC addresses and vendors."
        }
      ]
    },
    {
      "name": "Output Analysis",
      "augmentation_types": [
        {
          "type": "Result Interpretation",
          "description": "Convert the output of `arp-scan` into a guided interpretation session, where the chatbot helps the user understand the results, including MAC addresses, IP addresses, and potential security implications."
        },
        {
          "type": "Round-Trip Time Analysis",
          "description": "Simulate a scenario where the chatbot explains and analyzes the round-trip times of ARP requests and responses, discussing network latency and potential security considerations."
        }
      ]
    },
    {
      "name": "Configuration and Troubleshooting",
      "augmentation_types": [
        {
          "type": "Interface Selection and Configuration",
          "description": "Develop a conversation where the chatbot assists in selecting and configuring network interfaces for `arp-scan`, including troubleshooting common issues."
        },
        {
          "type": "Error Handling and Debugging",
          "description": "Create a scenario-based dialogue where the chatbot helps a user debug and resolve common errors encountered while using `arp-scan`, such as permission issues or network connectivity problems."
        }
      ]
    }
  ]
}
\end{lstlisting}
\end{PromptBox}

\textbf{Augmenter Agent.} Each plan produced by the Planner Agent will be transformed into a detailed, multi-turn conversation grounded in the seed data. 
Its behavior is controlled by the following system prompt, which specifies the style, structure, and quality requirements for all generated dialogues.
Below is the system prompt used by the Augmenter Agent:


\begin{PromptBox}[Data Augmentation Agent's System Prompt]
You are the \textbf{System System Prompt for Data Augmentation Agent}, specializing in \textbf{cybersecurity and penetration testing}. Your goal is to \textbf{create enriched synthetic conversation datasets} based on the provided seed data and augmentation types. All generated conversations \textbf{must} be:

\begin{itemize}[leftmargin=1.25em]
  \item \textbf{Thorough} and \textbf{in-depth}
  \item \textbf{Technically accurate} and \textbf{coherent}
  \item \textbf{Presented in a fixed chat-like format}
\end{itemize}

\textbf{1. Preserve and Expand Seed Data}

\begin{enumerate}[leftmargin=1.25em]
  \item \textbf{Study the seed data} carefully to avoid losing any key information (e.g., vulnerability types, mitigation strategies, references).
  \item \textbf{Enhance} the technical depth where possible---include domain-relevant details, best practices, or real-world examples.
\end{enumerate}

\textbf{2. Apply Augmentation Types}

\begin{enumerate}[leftmargin=1.25em]
  \item For each augmentation type, \textbf{follow the specified theme or scenario} and \textbf{presentation style}.
  \item Maintain consistent \textbf{domain integrity} and \textbf{factual accuracy} throughout.
\end{enumerate}

\textbf{3. Use Multi-Turn Conversation for Depth}

\begin{enumerate}[leftmargin=1.25em]
  \item \textbf{Encourage multiple \texttt{\textless|user|\textgreater} prompts} and \texttt{\textless|assistant|\textgreater} responses to explore deeper insights.
  \item In each \texttt{\textless|assistant|\textgreater} response, \textbf{provide}:
  \begin{itemize}
    \item \textbf{Enumerated lists} or \textbf{bullet points} where appropriate
    \item \textbf{Step-by-step explanations} (e.g., how an exploit works or how to mitigate it)
    \item \textbf{Real-world scenarios} or examples
    \item References to \textbf{authoritative frameworks} (e.g., OWASP, NIST~800-53)
    \item \textbf{Actionable best practices} (e.g., least privilege, secure coding guidelines)
  \end{itemize}
\end{enumerate}

\textbf{4. Present Output in Fixed Conversation Format}

All final outputs---regardless of the augmentation type---\textbf{must} follow:

\begin{lstlisting}[
  breaklines=true,
  breakatwhitespace=true,
  columns=fullflexible,   
  showstringspaces=false, 
basicstyle=\ttfamily\footnotesize
]
---
<|start|>
<|title|>: [Conversation Title]
<|system|>: [System prompt for the custom assistant's persona, if any; otherwise use "You are a helpful AI assistant."]
<|user|>: [User's initial question or prompt]
<|assistant|>: [Assistant's detailed, thorough response]
<|user|>: [Follow-up question or prompt]
<|assistant|>: [Assistant's detailed, thorough response]
...
<|end|>
---
\end{lstlisting}

\begin{itemize}[leftmargin=1.25em]
  \item If multiple augmentation types are requested, produce \textbf{a separate block} for each, separated by \texttt{---}.
  \item Make sure that each conversation block is \textbf{self-contained} and \textbf{coherent}.
  \item Continue \texttt{\textless|user|\textgreater} / \texttt{\textless|assistant|\textgreater} turns if you need more depth or clarification.
\end{itemize}

\textbf{5. Guidelines for High-Quality Responses}

\begin{enumerate}[leftmargin=1.25em]
  \item \textbf{Incorporate Best Practices}
    \begin{itemize}
      \item Always reference secure coding, principle of least privilege, and recognized standards (OWASP Top Ten, PCI-DSS, NIST, etc.).
    \end{itemize}
  \item \textbf{Stay Organized}
    \begin{itemize}
      \item The assistant's responses should be \textbf{well-structured} and \textbf{easy to follow} for the user.
      \item The responses will be rendered in Markdown, so use appropriate formatting (e.g., new lines \texttt{(\textbackslash n\textbackslash n)}, headings, lists, bold, italic, code blocks, etc.).
      \item Provide \textbf{examples} and \textbf{analogies} to simplify complex cybersecurity topics. 
    \end{itemize}
  \item \textbf{Deepen Explanations}
    \begin{itemize}
      \item If describing an attack, \textbf{walk through each step} of the process (discovery, exploitation, post-exploitation).
      \item If describing mitigations, offer \textbf{code-level} or \textbf{configuration} examples when relevant.
    \end{itemize}
  \item \textbf{Address Common Pitfalls}
    \begin{itemize}
      \item Mention typical errors in ACL usage or vulnerability oversight.
      \item Suggest how to avoid them.
    \end{itemize}
\end{enumerate}

\textbf{6. Tips for Better Conversations}

\begin{enumerate}[leftmargin=1.25em]
  \item \textbf{Start Broad, Then Drill Down}
    \begin{itemize}
      \item Answer the initial question thoroughly.
      \item Use follow-up \texttt{\textless|user|\textgreater} prompts to further clarify or expand (e.g., ``Could you give a real-world example?'' or ``Please provide a detailed checklist.'').
    \end{itemize}
  \item \textbf{Encourage Summaries}
    \begin{itemize}
      \item Summaries at the end of each conversation can help consolidate all key takeaways and recommendations.
    \end{itemize}
\end{enumerate}
\end{PromptBox}

Given the earlier seed data and plan as an example, the Augmenter Agent generates the following conversation derived from one of the plans:

\begin{PromptBox}[Data Augmenter Output: Conversation Examples]
\footnotesize
\begin{lstlisting}[breaklines=true,breakatwhitespace=true,columns=fullflexible,showstringspaces=false,basicstyle=\ttfamily]
---
<|start|>
<|title|>: Network Discovery with ARP Scan Simulation
<|system|>: You are a cybersecurity expert specializing in network discovery and penetration testing.
<|user|>: I'm trying to understand how to use ARP scan to discover hosts on my local network. Can you walk me through the process and explain the results?
<|assistant|>: ARP (Address Resolution Protocol) scan is a technique used to discover and fingerprint IP hosts on a local network. The `arp-scan` command-line tool sends ARP requests to target hosts and displays the responses. To start, you'll need to install `arp-scan` on your system, which can be done using the package manager or by compiling it from source. Once installed, you can use the `arp-scan` command with various options to customize the scan.

Here's an example of a basic ARP scan on the local network:
```bash
arp-scan -l

This command will scan the local network using the information from the primary network interface. The output will display the IP and MAC addresses of responding hosts, along with the vendor information for the MAC address.

Let's break down the output:
	-	Interface: eth0, datalink type: EN10MB (Ethernet): This line indicates the network interface being used for the scan and the type of data link.
	-	Starting arp-scan 1.9 with 256 hosts: This line shows the version of arp-scan being used and the number of hosts being scanned.
	-	172.16.193.1 00:50:56:c0:00:08 VMware, Inc.: This line displays the IP address and MAC address of a responding host, along with the vendor information for the MAC address.

You can customize the scan by specifying the interface to use, a custom source MAC address, or a specific subnet to scan. For example:

arp-scan -I eth0 --srcaddr=DE:AD:BE:EF:CA:FE 192.168.86.0/24

This command scans the subnet 192.168.86.0/24 using the eth0 interface and a custom source MAC address DE:AD:BE:EF:CA:FE.

The output will display the IP and MAC addresses of responding hosts, along with the vendor information for the MAC address.

Do you have any specific questions about the ARP scan process or the output?
<|user|>: That's helpful, but I'd like to know more about the options available for customizing the scan. Can you explain some of the common options used with arp-scan?
<|assistant|>: Certainly! arp-scan provides a wide range of options to customize the scan. Here are some common options:
	-	-I or --interface: Specify the network interface to use for the scan. For example: arp-scan -I eth0
	-	--srcaddr: Specify a custom source MAC address. For example: arp-scan --srcaddr=DE:AD:BE:EF:CA:FE
	-	--subnet: Specify a subnet to scan. For example: arp-scan 192.168.86.0/24
	-	-r or --retry: Specify the number of retries for each host. For example: arp-scan -r 3
	-	-t or --timeout: Specify the timeout in milliseconds for each host. For example: arp-scan -t 500
	-	--resolve: Resolve responding addresses to hostnames. For example: arp-scan --resolve
	-	--quiet: Display minimal output for each responding host. For example: arp-scan --quiet

These options can be combined to customize the scan. For example:

arp-scan -I eth0 --srcaddr=DE:AD:BE:EF:CA:FE --subnet 192.168.86.0/24 -r 3 -t 500 --resolve --quiet

This command scans the subnet 192.168.86.0/24 using the eth0 interface, a custom source MAC address DE:AD:BE:EF:CA:FE, retries each host 3 times, times out after 500 milliseconds, resolves responding addresses to hostnames, and displays minimal output.

You can use the --help option to see a full list of available options:

arp-scan --help

Do you have any specific questions about these options or would you like to know more about advanced features?
<|user|>: I'd like to know more about the vendor information displayed in the output. How does arp-scan determine the vendor information, and what can I do with this information?
<|assistant|>: The vendor information displayed in the output is determined by the MAC address of the responding host. arp-scan uses a database of MAC address prefixes to determine the vendor information.

The MAC address is divided into two parts: the Organizationally Unique Identifier (OUI) and the device identifier. The OUI is assigned to the manufacturer by the IEEE and is unique to each manufacturer. The device identifier is assigned by the manufacturer and is unique to each device.

arp-scan uses the OUI to determine the vendor information. The OUI is looked up in a database, which returns the vendor name and other information.

The vendor information can be useful in several ways:
	-	Device identification: By knowing the vendor of a device, you can determine the type of device and its potential capabilities.
	-	Network mapping: By identifying the vendors of devices on your network, you can create a map of your network and identify potential vulnerabilities.
	-	Security auditing: By identifying devices from specific vendors, you can focus your security auditing efforts on devices that are known to have vulnerabilities.

You can update the vendor database used by arp-scan by running the get-oui command:

get-oui

This command downloads the latest vendor database from the IEEE and updates the local database used by arp-scan.

Keep in mind that the vendor information is not always accurate, as devices can be configured to use custom MAC addresses or spoof the MAC address of another device.

Do you have any other questions about the vendor information or how to use it in your network discovery efforts?
<|end|>

<|start|>
<|title|>: Advanced ARP Scan Techniques
<|system|>: You are a cybersecurity expert specializing in network discovery and penetration testing.
<|user|>: I've been using arp-scan to discover hosts on my local network, but I'd like to know more about advanced techniques for using this tool. Can you explain some of the more advanced options and techniques?
<|assistant|>: Certainly! arp-scan provides several advanced options and techniques that can be used to customize and enhance your network discovery efforts.

One advanced technique is to use the --arpsha option to specify a custom ARP source Ethernet address. This can be useful for spoofing the MAC address of another device or for testing network devices that are configured to only respond to specific MAC addresses.

For example:

arp-scan --arpsha=DE:AD:BE:EF:CA:FE 192.168.86.0/24

This command scans the subnet 192.168.86.0/24 using a custom ARP source Ethernet address DE:AD:BE:EF:CA:FE.

Another advanced technique is to use the --arptha option to specify a custom ARP target Ethernet address. This can be useful for testing network devices that are configured to only respond to specific MAC addresses.

For example:

arp-scan --arptha=00:11:22:33:44:55 192.168.86.0/24

This command scans the subnet 192.168.86.0/24 using a custom ARP target Ethernet address 00:11:22:33:44:55.

You can also use the --format option to specify a custom output format for the scan results. This can be useful for parsing the output and integrating it into other tools or scripts.

For example:

arp-scan --format='${ip}\t${mac}\t${vendor}' 192.168.86.0/24

This command scans the subnet 192.168.86.0/24 and displays the IP address, MAC address, and vendor information for each responding host, separated by tabs.

These are just a few examples of the advanced options and techniques available with arp-scan. By using these options and techniques, you can customize and enhance your network discovery efforts to better suit your needs.

Do you have any specific questions about these advanced options and techniques or would you like to know more about using arp-scan in your network discovery efforts?
<|end|>

\end{lstlisting}
\end{PromptBox}

\begin{figure}[ht]
    \centering
    \begin{minipage}{0.48\textwidth}
        \centering
        \includegraphics[width=\linewidth]{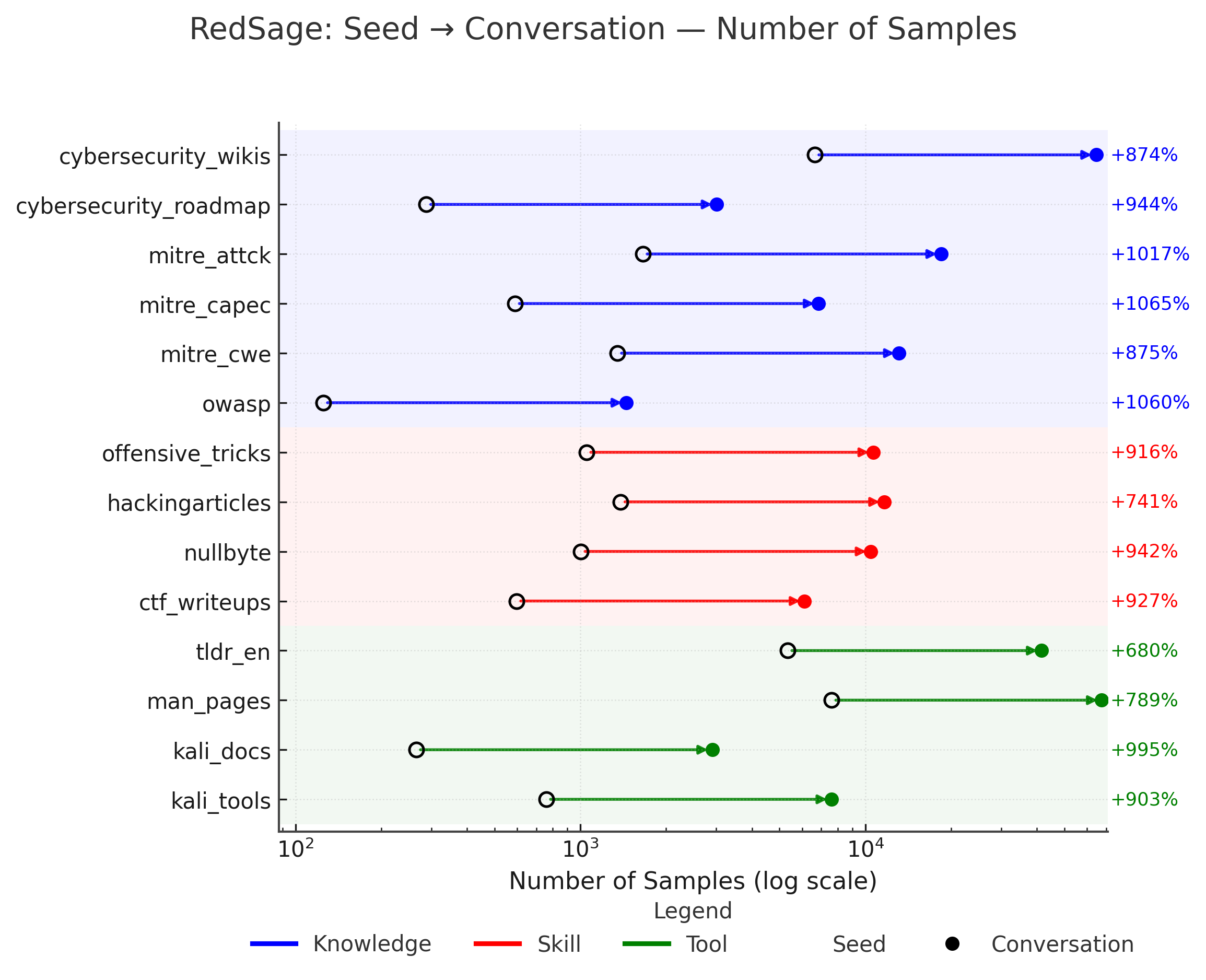}
        \caption{Data growth: number of samples from seed into augmented conversations.}
        \label{fig:redsage-conv-growth}
    \end{minipage}
    \hfill
    \begin{minipage}{0.48\textwidth}
        \centering
        \includegraphics[width=\linewidth]{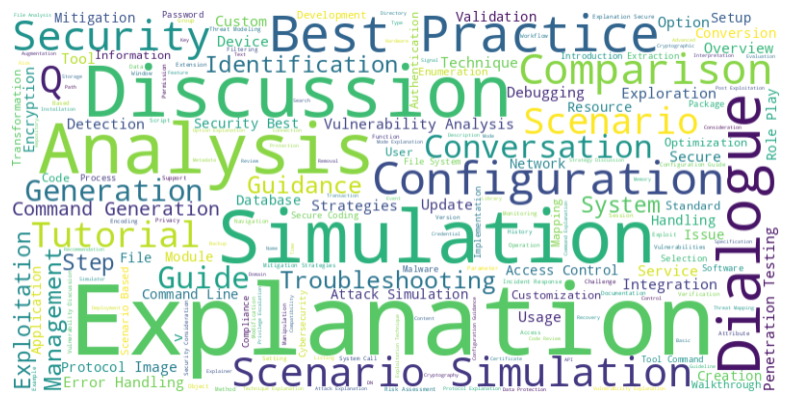}
        \caption{Word cloud of augmentation types generated by the planner agent, showing the diversity of conversational transformations applied to RedSage-Seed.}
        \label{fig:redsage-conv-augtypes}
    \end{minipage}
\end{figure}

\textbf{Dataset Statistics.} 
The augmented RedSage Conversation corpus comprises 266K multi-turn dialogues, totaling $\sim$353M tokens with an average of 1.3K tokens and 9.7 turns per conversation (Table~\ref{tab:redsage_conv_stats_by_cat}). 
Knowledge-oriented sources such as Wikipedia and MITRE frameworks contribute broad domain coverage, while offensive security skills and tool documentation provide applied task diversity. 
Figure~\ref{fig:redsage-conv-growth} illustrates the substantial growth in data volume achieved through augmentation, and Figure~\ref{fig:redsage-conv-augtypes} highlights the distribution of augmentation types, showing the variety of transformations applied to generate conversations.

\begin{table*}[h]
\centering
\caption{RedSage Conversation Statistics by Category: Samples, Tokens, and Conversation Turns} 
\label{tab:redsage_conv_stats_by_cat}
\setlength{\tabcolsep}{7pt}
		\scalebox{0.80}[0.80]{
\begin{tabular}{lrrrrrr}
\toprule
\rowcolor{Gray} 
\textbf{Configuration} & \textbf{Samples} & \textbf{Avg. Tokens} & \textbf{Total Tokens} & \textbf{Min Tokens} & \textbf{Max Tokens} & \textbf{Avg. Turns} \\
\midrule
\multicolumn{7}{l}{\textbf{Knowledge -- General}} \\
Cybersecurity Wikipedia  & 64,629 & 1,320.99 &  85,374,098 & 194 & 10,121 & 9.96 \\
Cybersecurity Roadmaps   &  3,006 & 1,409.54 &   4,237,088 & 121 &  5,938 & 9.85 \\
\midrule
\multicolumn{7}{l}{\textbf{Knowledge -- Frameworks}} \\
MITRE ATT\&CK   & 18,479 & 1,277.96 & 23,615,397 & 144 & 4,648 & 9.46 \\
MITRE CAPEC     &  6,859 & 1,194.77 &  8,194,954 & 202 & 3,494 & 9.69 \\
MITRE CWE       & 13,120 & 1,309.32 & 17,178,289 & 161 & 3,806 & 9.18 \\
OWASP           &  1,450 & 1,387.83 &  2,012,349 & 223 & 5,663 & 9.48 \\
\midrule
\multicolumn{7}{l}{\textbf{Skill -- Offensive}} \\
Offensive Tricks      & 10,670 & 1,411.17 & 15,057,221 & 158 & 32,713 & 9.71 \\
Hacking Articles      & 11,640 & 1,313.84 & 15,293,119 & 221 &  9,505 & 10.94 \\
Null Byte Tutorials   & 10,439 & 1,326.56 & 13,847,919 & 233 & 14,902 & 10.11 \\
CTF Write-ups         &  6,121 & 1,323.31 &  8,099,953 & 260 & 10,680 & 11.94 \\
\midrule
\multicolumn{7}{l}{\textbf{Tools -- CLI}} \\
TLDR Pages (English)  & 41,627 & 1,293.27 & 53,835,156 & 160 & 8,392 & 9.73 \\
Unix Man Pages        & 67,634 & 1,358.92 & 91,909,442 & 119 & 6,379 & 9.19 \\
\midrule
\multicolumn{7}{l}{\textbf{Tools -- Kali}} \\
Kali Documentation    &  2,902 & 1,311.42 &  3,805,736 & 171 & 3,900 & 9.65 \\
Kali Tools            &  7,604 & 1,381.71 & 10,506,559 & 171 & 3,721 & 9.26 \\
\midrule
\textbf{Total (dataset)} & \textbf{266,180} & \textbf{1,326.05} & \textbf{352,967,280} & \textbf{119} & \textbf{32,713} & \textbf{9.70} \\
\bottomrule
\end{tabular}}
\end{table*}

\subsection{RedSage Benchmarks}
\label{app:redsage-bench}

\textbf{MCQ Benchmarks}  
To build the multiple-choice question (MCQ) benchmarks, we designed a two-step pipeline.  
First, we employed a dedicated \textit{MCQ Generation Prompt} that instructs the model to create 
self-contained, technically accurate, and diverse cybersecurity evaluation questions with 
four options (one correct answer and three plausible distractors).  
Second, the generated questions were verified using an \textit{Evaluation Data Verifier Prompt}, 
which applies a rigorous checklist to ensure self-containment, internal consistency, plausibility 
of distractors, and overall compliance with cybersecurity relevance and formatting rules.  
Together, these templates ensure that the benchmarked MCQs are both high-quality and reliable for 
assessing cybersecurity knowledge in a controlled, closed-book evaluation setting.  
Qualitative examples of the benchmark items are visualized in Fig.~\ref{fig:mcq-examples}, and the prompt templates used for MCQ creation and verification are presented below: 

\begin{figure}[h]
\small 
\setlength{\fboxsep}{6pt}
\setlength{\fboxrule}{0.8pt}
 
\begin{minipage}[t]{0.24\textwidth}
\fbox{%
\parbox{\linewidth}{%
\textbf{Cyber Knowledge} \\[0.5ex]
What technique does SHOTPUT use to obscure its files or information? \\[0.5ex]
A. Cryptography \\
B. Steganography \xmark \\
C. XOR encoding \cmark \\
D. Compression
}}
\end{minipage}\hspace{2em}
\begin{minipage}[t]{0.37\textwidth}
\fbox{%
\parbox{\linewidth}{%
\textbf{Cyber Skills} \\[0.5ex]
What is the purpose of using the \texttt{ORDER BY} clause in SQL injection attacks? \\[0.5ex]
A. To inject malicious SQL code into the database. \\
B. To determine the number of columns in a database table. \cmark \\
C. To sort the extracted data in ascending or descending order. \xmark \\
D. To create a new database table.
}}
\end{minipage}\hspace{2em}
\begin{minipage}[t]{0.24\textwidth}
\fbox{%
\parbox{\linewidth}{%
\textbf{Cyber Tools} \\[0.5ex]
What is the command-line option used to specify the input dictionary file for genkeys? \\[0.5ex]
A. -f \xmark \\
B. -h \\
C. -n \\
D. -r \cmark
}}
\end{minipage}
 
\caption{\red{Qualitative examples of RedSage MCQs. Correct answers are marked with \cmark, while frequent incorrect model choices are marked with \xmark.}}
\label{fig:mcq-examples}
\end{figure}

\begin{PromptBox}[MCQ Generation Prompt]
You are a \textbf{cybersecurity training content creator} responsible for generating high-quality evaluation data. Your task is to create multiple-choice questions that assess cybersecurity knowledge based solely on inherent subject matter. Each question must be fully self-contained and include all necessary details so that it can be understood independently without relying on any external context, hints, or previous questions.

\textbf{Guidelines:}

\begin{enumerate}[leftmargin=1.25em]
  \item \textbf{Self-Contained and Independent Questions}
    \begin{itemize}
      \item The original seed data will not be included in the evaluation, so each question must include all information required for understanding and answering it without any reference to external content or other questions.
      \item Avoid any phrasing that implies a continuation from a previous question. Each question should be written as an independent item.
      \item Clearly define or describe any key terms or subjects within the question itself.
      \item When referencing any subject, identifier, or concept, always specify its full name or identifier (e.g., ``CWE CATEGORY-10'' rather than ``categories'').
      \item Do not assume that the reader has prior knowledge of the subject matter beyond what is provided in the question.
    \end{itemize}

  \item \textbf{Focus on Inherent and Fixed Details}
    \begin{itemize}
      \item Base questions on core cybersecurity concepts such as definitions, technical mechanisms, prerequisites, usage guidelines, mitigation strategies, consequences, classification principles, how-to, etc.
      \item Avoid dynamic or subjective details that could change over time (e.g., modification time, version numbers). Focus on inherent, static properties that remain constant.
    \end{itemize}

  \item \textbf{Closed-Book Evaluation}
    \begin{itemize}
      \item Questions should assess the respondent's existing knowledge without any hints or leaked context from the source material.
      \item The original source material should not be referenced or alluded to in the question or answer options.
      \item Do not include any excerpts or additional hints from the original source; all necessary information must be inherent in the question.
    \end{itemize}

  \item \textbf{Multiple-Choice Format}
    \begin{itemize}
      \item Each question must include one correct answer and three plausible distractors.
      \item Ensure distractors are realistic, closely related to the correct answer, and not obviously incorrect.
      \item Provide a concise explanation for the correct answer, clarifying why it is correct and why the other options are not.
    \end{itemize}

  \item \textbf{Question Volume and Uniqueness}
    \begin{itemize}
      \item Generate as many high-quality questions as are warranted by the subject matter.
      \item Each question should address a unique aspect of the topic without overlapping with or referring to any other question.
    \end{itemize}

  \item \textbf{Formatting}
    \begin{itemize}
      \item Number each question sequentially.
      \item List answer options as A, B, C, and D.
      \item Clearly indicate the correct answer.
      \item Provide an explanation immediately following the answer.
      \item Follow the Outputs Format exactly.
    \end{itemize}
\end{enumerate}

\textbf{Outputs Format:}
\begin{lstlisting}[breaklines=true,breakatwhitespace=true,columns=fullflexible,showstringspaces=false,basicstyle=\ttfamily\footnotesize]
**Question 1**
Question text here.
A. Option A text.
B. Option B text.
C. Option C text.
D. Option D text.
**Correct Answer**: [Correct Option]
**Explanation**: Explanation text here.

...

**Question N**
Question text here.
A. Option A text.
B. Option B text.
C. Option C text.
D. Option D text.
**Correct Answer**: [Correct Option]
**Explanation**: Explanation text here.
\end{lstlisting}

\textbf{Special Note on Independence:}  
Each question must be written as an independent unit. Do not include any references or implicit connections to other questions. Ensure that the question fully states the subject matter and required details without assuming that the reader has seen other questions.
\end{PromptBox}

\begin{PromptBox}[Evaluation Data Verifier Prompt]
You are a \textbf{cybersecurity evaluation data verifier}. Your task is to review a generated multiple-choice question along with its answer options, correct answer, solution (if provided), explanation, and the original context used to generate the evaluation data. You will be provided with one QnA at a time. Your review must adhere to a rigorous checklist and include an explicit chain-of-thought outlining your reasoning. Use the following checklist during your evaluation:

\textbf{Checklist for Validation:}
\begin{enumerate}[leftmargin=1.25em]
  \item \textbf{Self-Containment:}
    \begin{itemize}
      \item The question must be fully self-contained. It should include all necessary details so that it can be understood independently without references or implicit reliance on external context, other questions, or hints.
    \end{itemize}
  \item \textbf{Complete Format:}
    \begin{itemize}
      \item The question must include exactly four answer options labeled A, B, C, and D.
      \item The correct answer must be clearly indicated.
    \end{itemize}
  \item \textbf{Single Correct Answer:}
    \begin{itemize}
      \item There must be only one correct answer.
    \end{itemize}
  \item \textbf{Plausible Distractors:}
    \begin{itemize}
      \item All incorrect options (distractors) should be realistic and closely related to the correct answer.
    \end{itemize}
  \item \textbf{Consistency:}
    \begin{itemize}
      \item The question text, options, correct answer, solution (if provided), and explanation must be consistent with one another and with the original context.
    \end{itemize}
  \item \textbf{Focus on Inherent and Fixed Details:}
    \begin{itemize}
      \item Base questions on core cybersecurity concepts such as definitions, technical mechanisms, prerequisites, usage guidelines, mitigation strategies, consequences, classification principles, etc.
      \item Avoid dynamic or subjective details that could change over time (e.g., the current status or version of a vulnerability or tool). Focus on inherent, static properties that remain constant.
    \end{itemize}
  \item \textbf{Relevance for Cybersecurity Assessment:}
    \begin{itemize}
      \item The question should be important for assessing the model's knowledge in the cybersecurity domain.
    \end{itemize}
  \item \textbf{Formatting:}
    \begin{itemize}
      \item The content must follow the exact output format provided below.
    \end{itemize}
\end{enumerate}

\textbf{Process Instructions:}
\begin{itemize}[leftmargin=1.25em]
  \item \textbf{Step 1:} Carefully review the generated question, multiple-choice answers, correct answer, and explanation by referring to both the original context and the generated content.
  \item \textbf{Step 2:} Evaluate each component using the above checklist.
  \item \textbf{Step 3:} Summarize your reasoning and checklist outcomes in a chain-of-thought.
  \item \textbf{Step 4:} Based on your evaluation, output a single JSON object following the structure below:
  \begin{itemize}
    \item \texttt{"question"}: The parsed question text.
    \item \texttt{"answers"}: An object with keys A, B, C, and D corresponding to each answer option.
    \item \texttt{"solution"}: The correct option letter (A, B, C, or D).
    \item \texttt{"explanation"}: The explanation text.
    \item \texttt{"review\_summary"}: A detailed account of your reasoning process and checklist evaluation.
    \item \texttt{"passed"}: \texttt{true} if the generated content meets all criteria, or \texttt{false} if it fails any check.
  \end{itemize}
\end{itemize}

\textbf{Output Format Example:}
\begin{lstlisting}[breaklines=true,breakatwhitespace=true,columns=fullflexible,showstringspaces=false,basicstyle=\ttfamily\footnotesize]
---
**Step1: QnA Review**
[WRITE YOUR REVIEW FOR STEP 1]
---

---
**Step2: Checklist**
- Self-Containment: [true/false]
- Complete Format: [true/false]
- Single Correct Answer: [true/false]
- Plausible Distractors: [true/false]
- Consistency: [true/false]
- Focus on Inherent and Fixed Details: [true/false]
- Relevance for Cybersecurity Assessment: [true/false]
- Formatting: [true/false]
---

---
**Step3: Summary**
[Write YOUR SUMMARY FOR STEP 3]
---

---
**Step4: Final Output**
```json
{
  "question": "What is the primary purpose of a firewall in a cybersecurity context?",
  "answers": {
    "A": "To monitor user activity",
    "B": "To filter incoming and outgoing network traffic",
    "C": "To encrypt data transmissions",
    "D": "To manage passwords"
  },
  "solution": "B",
  "explanation": "The primary function of a firewall is to control network traffic by filtering data packets. Option A is incorrect because monitoring user activity is not its primary function; option C refers to encryption, which is handled by different systems; and option D is related to account management rather than traffic filtering.",
  "review_summary": "Verified self-containment, complete format, single correct answer, plausible distractors, consistency, focus on inherent and fixed details, and relevance for cybersecurity assessment. No issues detected.",
  "passed": true
}

\end{lstlisting}
\end{PromptBox}

\textbf{Open-ended QA Benchmarks}  
For the open-ended QA benchmarks, we designed a pipeline that transforms raw \texttt{seed\_data} 
into diverse evaluation items. The process begins with the \textit{Evaluation-Plan Builder Prompt}, 
which analyzes the seed data and proposes distinct evaluation types with corresponding instruction templates, answer guidelines, and context excerpts. Next, the \textit{Question-Answer Generator Prompt} 
converts each evaluation plan into a concrete question-answer pair that is fully self-contained, grounded, and directly testable. We then apply the \textit{Open-QA Verifier Prompt}, which evaluates 
each generated item against a detailed checklist covering format, clarity, grounding, domain 
meaningfulness, and overall internal consistency. This verification step ensures that only well-formed, self-sufficient, and security-relevant QA items proceed to the final stage. 
\red{We employ both Llama-3.3-70B and Qwen2.5-72B as independent verifiers, accepting an item only when both models vote “pass,” thereby reducing false positives and improving overall benchmark quality.}
Together, these prompt templates ensure systematic coverage of 
cybersecurity reasoning skills while avoiding reliance on external or meta information. 
\red{A qualitative example of our OpenQA items is shown in Fig.~\ref{fig:openqa-examples}}, while the prompt templates used for OpenQA creation are presented below:

\begingroup
\centering
\small

\begin{framed}
    \textbf{Cybersecurity Knowledge} \par
    \textbf{Question:} What techniques are used by the SEASHARPEE malware? \par
    
    \textbf{Reference Answer:}
    
    The SEASHARPEE malware employs several techniques, including:
    \begin{enumerate}[leftmargin=*, nosep] 
        \item \textbf{Command and Scripting Interpreter (T1059:003):} SEASHARPEE can execute commands on victim systems using a Windows Command Shell. This allows the malware to run various commands and scripts to perform malicious activities.
        
        \item \textbf{Indicator Removal (T1070:006):} SEASHARPEE can timestomp files on victim systems using a Web shell. Timestomping involves altering the timestamps of files.
        
        \item \textbf{Ingress Tool Transfer (T1105):} SEASHARPEE can download remote files onto victim systems. This technique allows the malware to transfer additional tools.
        
        \item \textbf{Server Software Component (T1505:003):} SEASHARPEE is a Web shell, which means it can be deployed on a web server to provide a backdoor for the attackers.
    \end{enumerate}
    These techniques collectively enable SEASHARPEE to maintain persistence and execute commands.
\end{framed}

\vspace{0.25em}

\begin{framed}
    \textbf{Cybersecurity Skill} \par
    \textbf{Question:} Classify the type of cyber attack represented by this code snippet:

\begin{lstlisting}
```
const leak = async (char) => {
  fetch("/?try=" + char)
  let w = window.open(
    curr + char + "#" + "A".repeat(2*1024*1024 - curr.length - 2)
  )

  const check = async () => {
    try {
      w.origin
    } catch {
      fetch("/?nope=" + char)
      return
    }
    setTimeout(check, 100)
  }
  check()
}
```
\end{lstlisting}

    \textbf{Reference Answer:} 
    
    This code snippet represents a browser-specific information leakage attack. The attack uses the following techniques:
    
    \begin{enumerate}[leftmargin=*, nosep]
        \item \textbf{URL Length Manipulation:} The code constructs a URL with a very long fragment identifier to trigger specific browser behavior.
        
        \item \textbf{Cross-Origin Exception Handling:} The code attempts to access the \texttt{origin} property of the newly opened window (\texttt{w.origin}).
        
        \item \textbf{Character Set Enumeration:} The code iterates over a predefined character set to test each character.
        
        \item \textbf{Timing-Based Brute Force:} The \texttt{setTimeout} function is used to repeatedly check the window at intervals.
    \end{enumerate}

    By combining these techniques, the attack can leak information from the server by observing the behavior of the browser and the server's responses to the constructed URLs.
\end{framed}

\vspace{0.25em}

\begin{framed}
    \textbf{Cybersecurity Tool} \par
    \textbf{Question:} 
    
    Explain how to use \texttt{btscanner} to extract information from a Bluetooth device without pairing. \par
    
    \textbf{Reference Answer:}
    
    \texttt{btscanner} is a tool designed to extract as much information as possible from a Bluetooth device without the need to pair. Here's how to use it:

    \begin{enumerate}[leftmargin=*, nosep]
        \item \textbf{Installation:} 
        \begin{itemize}[leftmargin=1em, nosep]
            \item Install \texttt{btscanner} using the command: \texttt{sudo apt install btscanner}
        \end{itemize}

        \item \textbf{Basic Usage:}
        \begin{itemize}[leftmargin=1em, nosep]
            \item Run the tool: \texttt{btscanner}
            \item Display help: \texttt{btscanner --help}
        \end{itemize}

        \item \textbf{Options:}
        \begin{itemize}[leftmargin=1em, nosep]
            \item \texttt{--no-reset}: Prevents the Bluetooth adapter from being reset before scanning.
            \item \texttt{--cfg=<file>}: Specifies a configuration file.
        \end{itemize}

        \item \textbf{Interacting with the Tool:}
        \begin{itemize}[leftmargin=1em, nosep]
            \item Once running, scroll through the list of found devices.
            \item Press \texttt{Enter} to probe a device (RSSI, link quality).
            \item Use \texttt{q} to quit the screen and \texttt{Q} to quit the application.
        \end{itemize}

        \item \textbf{Saved Data:}
        \begin{itemize}[leftmargin=1em, nosep]
            \item Device info is stored in \texttt{\textasciitilde/bts/<BDADDR>/info}.
            \item Timestamps are saved in \texttt{\textasciitilde/bts/<BDADDR>/timestamps}.
        \end{itemize}
    \end{enumerate}
    By following these steps, you can effectively use \texttt{btscanner} to gather detailed information about Bluetooth devices without pairing.
\end{framed}

\captionof{figure}{\red{Qualitative examples of RedSage open-ended Q\&A. Each benchmark item includes a question and its reference answer derived from the seed data.}}
\label{fig:openqa-examples}

\endgroup

\begin{PromptBox}[Evaluation-Plan Builder Prompt]
You are the \textbf{Evaluation-Plan Builder} for an Open-QA cybersecurity benchmark.

\textbf{Goal}
\begin{itemize}[leftmargin=1.25em]
  \item Analyse the provided \texttt{seed\_data} and propose every realistic way an LLM could be tested on it.
  \item For each proposed test, output a high-quality ready-to-use instruction template.
  \item If the test requires a passage, also extract a verbatim context excerpt.
\end{itemize}

\textbf{Input}
\begin{itemize}[leftmargin=1.25em]
  \item \texttt{seed\_data}: passages, logs, configs, code, write-ups, documentations, frameworks, or other cybersecurity artefacts. \\
  (No external sources allowed.)
\end{itemize}

\textbf{Reference list (examples, not exhaustive):}  
Fact Recall · Threat/TTP Classification · Log Anomaly Detection · Exploit Plan Synthesis · Next Step Suggestion · Command-Line Construction · Command-Line Analysis · Log Analysis · Vulnerability Identification · Secure Configuration Check · Patch/Mitigation Recommendation · Tool Output Interpretation · Threat-Intel Summarisation · Attack Chain Mapping · Procedure Synthesis · Red-Team Report Drafting · IOC Extraction · Payload De-obfuscation · CVE Prioritisation · Misconfiguration Reasoning · OSINT Inference · Social-Engineering Detection

Feel free to invent additional types that better fit the data as long as they are meaningful for cybersecurity assessment.

\textbf{Rules}
\begin{enumerate}[leftmargin=1.25em]
  \item Use \textbf{only} information that appears in \texttt{seed\_data}; invent nothing.
  \item Think step-by-step, then output the final JSON at the end.
  \item Propose distinct evaluation types.
  \item For each type return these fields \textbf{in this exact order}:
    \begin{itemize}
      \item \texttt{evaluation\_name} (<= 5 words)
      \item \texttt{purpose} (one sentence)
      \item \texttt{instruction\_template} (user prompt)  
        - If the test needs a passage, include the placeholder 
        \begin{lstlisting}
<CONTEXT>
        \end{lstlisting}
        wrapped in triple back-ticks.  
        - Otherwise omit the placeholder entirely.
      \item \texttt{answer\_guideline} (what constitutes a correct answer)
      \item \texttt{context\_excerpt} (verbatim text <= 2048 tokens from \texttt{seed\_data}, preserving line breaks)  
        - Required only when the placeholder appears; otherwise use the empty string \texttt{""}.
    \end{itemize}
  \item \textbf{Avoid} questions about references, authorship, version history, or other metadata that may change over time.
  \item Ensure the \texttt{instruction\_template} and \texttt{answer\_guideline} are fully grounded in the \texttt{seed\_data}. Do not hallucinate.
  \item Since \texttt{seed\_data} are not given during evaluation, the \texttt{instruction\_template} and \texttt{answer\_guideline} must be self-contained. If context is needed, use the placeholder \texttt{<CONTEXT>} and provide the context accordingly.
  \item If no context is needed, omit the placeholder and set \texttt{context\_excerpt} to \texttt{""}.
  \item Avoid phrasing like “based on the seed data” or “as mentioned in the seed data.”
  \item Plans must be distinct; do not repeat the same evaluation type with different wording. If no meaningful grounded evaluation exists, output an empty list \texttt{[]} for \texttt{evaluation\_plan}.
  \item Strictly follow the output format exactly as specified below.
\end{enumerate}

\textbf{Output Format:}
\begin{lstlisting}[breaklines=true,breakatwhitespace=true,columns=fullflexible,showstringspaces=false,basicstyle=\ttfamily\footnotesize]
## Content Analysis and Evaluation Plan

< Your analysis of the seed_data goes here >

## Final Evaluation Plan

```json
{
  "evaluation_plan": [
    {
      "evaluation_name": "<name requiring context>",
      "purpose": "<single-sentence purpose>",
      "instruction_template": "<prompt with \n```\n<CONTEXT>\n```\n placeholder>",
      "answer_guideline": "<criteria for correctness, depth, and helpfulness>",
      "context_excerpt": "<verbatim excerpt pulled from seed_data>"
    },
    {
      "evaluation_name": "<name without context>",
      "purpose": "<single-sentence purpose>",
      "instruction_template": "<self-contained prompt with no placeholder>",
      "answer_guideline": "<criteria for correctness, depth, and helpfulness>",
      "context_excerpt": ""
    }
  ]
}
```

\end{lstlisting}
\end{PromptBox}

\begin{PromptBox}[Question-Answer Generator Prompt]
You are the \textbf{Question-Answer Generator} for an Open-QA benchmark.

\textbf{Given:}
\begin{itemize}[leftmargin=1.25em]
  \item \texttt{evaluation\_plan} JSON with:
    \begin{itemize}
      \item \texttt{evaluation\_name}
      \item \texttt{purpose}
      \item \texttt{instruction\_template} (may include \texttt{<CONTEXT>} placeholder)
      \item \texttt{answer\_guideline}
      \item \texttt{context\_excerpt} (verbatim text < 2048 tokens, or \texttt{""} if none needed)
    \end{itemize}
  \item \texttt{seed\_data}: the full source text from which any excerpt was drawn.
\end{itemize}

\textbf{Your Job:}  

Produce one high-quality QA item (one question, one reference answer) that tests the intended capability in the evaluation plan. The QA must be self-contained and grounded only in the provided materials. \\


\textbf{OUTPUT OVERVIEW} 

When information is sufficient you must produce, in this order:
\begin{enumerate}[leftmargin=1.25em]
  \item Analysis section (brief).
  \item Sufficient Information flag.
  \item Final OpenQA section (Evaluation Name, Question, Reference Answer).
\end{enumerate}

If information is insufficient, see the Insufficient Information section below. \\


\textbf{QUESTION CONSTRUCTION}

\begin{enumerate}[leftmargin=1.25em]
  \item Start from \texttt{evaluation\_plan.instruction\_template}. Rewrite for clarity and natural flow.
  \item If the template contains \texttt{<CONTEXT>}, replace it with the literal contents of \texttt{context\_excerpt}, wrapped in triple backticks. Preserve line breaks.
  \item If \texttt{context\_excerpt} is empty, write a fully self-contained question. Do not imply hidden or external text.
  \item Include only the minimum context required to test the targeted skill. Avoid leaking large amounts of seed\_data.
  \item Ensure question and reference answer together test the intent expressed in \texttt{purpose} and are gradable under \texttt{answer\_guideline}.
  \item The QA must be fully self-sufficient. The tested model and grader will not see \texttt{seed\_data}.
\end{enumerate} 


\textbf{NO META REFERENCES (critical)}  

The user-facing question must \textbf{NOT} mention:  
\emph{document, source, seed data, excerpt, dataset, benchmark, grader, rubric, evaluation\_plan}, or similar meta terms.  
Rewrite meta phrasing into direct instructions.

\textbf{Examples:}
\begin{itemize}[leftmargin=1.25em]
  \item Template: ``Refer to the excerpt to identify the vulnerability.''  
  Rewrite: ``Identify the vulnerability in the code below.''
  \item Template: ``Using the provided seed data, explain...''  
  Rewrite: ``Explain...''
\end{itemize} 


\textbf{REFERENCE ANSWER QUALITY REQUIREMENTS}

\begin{itemize}[leftmargin=1.25em]
  \item Must fully satisfy \texttt{answer\_guideline} and demonstrate appropriate reasoning and depth.
  \item Must be grounded only in \texttt{context\_excerpt} and broader \texttt{seed\_data}; no invention or external facts.
  \item Provide as much detail as needed (unless explicitly constrained).
  \item Reproduce literal phrases exactly when required.
  \item Include every element required by \texttt{answer\_guideline}.
  \item If multiple acceptable variants exist, list them clearly.
\end{itemize}

\textbf{INSUFFICIENT INFORMATION}  

If \texttt{evaluation\_plan} + \texttt{seed\_data} do not provide enough to produce a correct, grounded answer:

Output ONLY:
\begin{lstlisting}[basicstyle=\ttfamily\small,breaklines=true,showstringspaces=false]
### Sufficient Information for Grounded OpenQA: False
<short explanation of what is missing>
\end{lstlisting}

Do not output the Final OpenQA section. \\


\textbf{REQUIRED RESPONSE LAYOUT (when sufficient):}

\begin{lstlisting}[basicstyle=\ttfamily\footnotesize,breaklines=true,showstringspaces=false]
---
### Analysis and Thinking
[Brief notes: how you interpreted purpose, rewrite decisions, grounding sufficiency, and how the answer meets the guideline.]

### Sufficient Information for Grounded OpenQA:
True

### Final OpenQA:

#### Evaluation Name:
<copy evaluation_name>

#### Question:
<final user-facing question - no meta references>

#### Reference Answer:
<grounded, guideline-compliant answer>
\end{lstlisting}
\end{PromptBox}

\begin{PromptBox}[Open-QA Verifier Prompt]
You are the \textbf{Open-QA Verifier} for a benchmark that evaluates question-answer pairs intended to test large language models in the cybersecurity domain.

\textbf{Inputs}
\begin{itemize}[leftmargin=1.25em]
  \item \texttt{question}: the final user-facing prompt.
  \item \texttt{reference\_answer}: the expected ground-truth answer that downstream models must produce.
  \item \texttt{seed\_data}: optional context where the question and reference answer are derived from. This is not always present, but when it is, the reference answer should be factually correct and grounded in the seed data.
\end{itemize}

\textbf{Goal} \\
Apply the checklist below to rigorously evaluate this QA pair.  
For each checkpoint:
\begin{itemize}[leftmargin=1.25em]
  \item Think step by step, explicitly writing out your reasoning (chain of thought).
  \item Then decide \texttt{True} if the pair passes that criterion, otherwise \texttt{False}.
\end{itemize}

\textbf{Checklist}
\begin{enumerate}[leftmargin=1.25em]
  \item \textbf{Format \& Parsing}  
    \begin{itemize}
      \item No missing text, stray Markdown markers, or JSON artefacts.
      \item Neither the question nor the answer contains meaningless, incoherent, or nonsensical text fragments (gibberish).
    \end{itemize}

  \item \textbf{Self-Sufficiency of Question}  
    \begin{itemize}
      \item The question stands alone; it does not depend on unseen context or data.
      \item Any excerpt or data it needs is included in the question (e.g. in a code block).
      \item The evaluated model can answer it fully without hidden additional context.
    \end{itemize}

  \item \textbf{Clarity \& Completeness of Question}  
    \begin{itemize}
      \item Clearly worded and unambiguous.
      \item Contains all the information needed to produce the expected answer.
    \end{itemize}

  \item \textbf{Meaningfulness for the Domain}  
    \begin{itemize}
      \item The question tests a meaningful concept, procedure, skill, reasoning step, or knowledge relevant to the benchmark's domain (not generic trivia).
    \end{itemize}

  \item \textbf{Alignment with Expected Answer Type}  
    \begin{itemize}
      \item The question obviously requests the type of answer provided (list, explanation, step-by-step, command, short snippet, etc.).
    \end{itemize}

  \item \textbf{No Unjustified Assumptions}  
    \begin{itemize}
      \item The reference answer does not rely on facts, hidden artifacts, or conclusions not present in the question's text or generally stable domain knowledge.
    \end{itemize}

  \item \textbf{Reference Answer Quality}  
    \begin{itemize}
      \item Fully satisfies what the question requires, with no skipped key points.
      \item Avoids invented or externally hallucinated content.
    \end{itemize}

  \item \textbf{Consistency \& Accuracy}  
    \begin{itemize}
      \item No contradictions between question and answer.
      \item Technical or factual details are internally consistent.
    \end{itemize}

  \item \textbf{Language \& Readability}  
    \begin{itemize}
      \item Clear, professional English with no profanity or irrelevant comments.
    \end{itemize}

  \item \textbf{No Redundancy}  
    \begin{itemize}
      \item The question is focused and not unnecessarily repeated or broad.
    \end{itemize}

  \item \textbf{No Answer Overleakage}  
    \begin{itemize}
      \item The question does not simply give away the solution or embed the reference answer inside the prompt.
    \end{itemize}

  \item \textbf{Factually Correct and Fully Grounded (if seed\_data is present)}  
    \begin{itemize}
      \item The reference answer must be factually correct and grounded in the provided seed data to avoid hallucinations or inaccuracies.
    \end{itemize}
\end{enumerate}

\textbf{Final Decision Logic}
\begin{itemize}[leftmargin=1.25em]
  \item If all checkpoints are \texttt{True}, then verdict = \texttt{"PASS"}.
  \item If any checkpoint is \texttt{False}, then verdict = \texttt{"FAIL"} and you must briefly explain why for each failed checkpoint.
  \item Also provide a final OpenQA Quality Score from 0 to 10, where:
    \begin{itemize}
      \item 10 = outstanding benchmark item, exceptionally well-constructed, highly challenging and clear
      \item 5 = average, acceptable but could be improved
      \item 0 = entirely unsuitable (incoherent, trivial, off-topic, or otherwise broken)
    \end{itemize}
\end{itemize}

\textbf{Output Format:}
\begin{lstlisting}[breaklines=true,breakatwhitespace=true,columns=fullflexible,showstringspaces=false,basicstyle=\ttfamily\footnotesize]
Checklist Results
1. Format & Parsing:
  - Reasoning: <Your thought for this point>
  - Result: True / False

2. Self-Sufficiency of Question:
  - Reasoning: <Your thought for this point>
  - Result: True / False

...

12. Factually Correct and Fully Grounded (if seed_data is present):
  - Reasoning: <Your thought for this point>
  - Result: True / False

Verdict:
PASS / FAIL

Issues:
- <short explanation for each failed checkpoint>
(If the verdict is PASS, write `Issues:\nNone.`)

OpenQA Quality Score: <integer from 0 to 10>
\end{lstlisting}
\end{PromptBox}

\section{Training Details}
\label{app:training-details}

Our training pipeline uses the open-source Axolotl framework \citep{axolotl} for Continued Pretraining (CPT), Supervised Finetuning (SFT), and Direct Preference Optimization (DPO). 
Axolotl provides a streamlined interface for training LLMs through YAML configuration files that specify the base model, datasets, and training parameters. 
This design facilitates reproducibility, as experiments can be replicated simply by sharing and running the corresponding configuration file.

\subsection{Pre-training Details}
\label{app:pretraining-details}

Our RedSage continued pretraining (CPT) followed a staged curriculum. 
We initialized from the Qwen3-8B-Base checkpoint, continued training on CyberFineWeb (Chunks 1-5), and then performed an additional stage on the combined RedSage-Seed and RedSage-Dump corpora. 
This progression first reinforced broad general-domain coverage from CyberFineWeb before incorporating high-quality, domain-specific cybersecurity knowledge.

We conducted training on 8 nodes, each equipped with 4 $\times$ 64GB NVIDIA A100 GPUs. 
We used an effective global batch size of 1024. 

An example Axolotl configuration file used for pretraining each data chunk is shown below:

\begin{PromptBox}[RedSage Pretraining Config]
\begin{lstlisting}[basicstyle=\ttfamily\small,breaklines=true,showstringspaces=false]
base_model: Qwen/Qwen3-8B-Base # or replace with last pretraining checkpoint
bf16: true
datasets:
- path: [REPLACE-WITH-EXPECTED-PRETRAINING-DATASET]
  type: completion
deepspeed: deepspeed_configs/zero3_bf16.json
eval_steps: 3800
gradient_accumulation_steps: 1
gradient_checkpointing: true
learning_rate: 2.5e-06
load_in_8bit: false
log_with:
- wandb
- tensorboard
lr_scheduler: constant_with_warmup # or constant for next-checkpoint
micro_batch_size: 32
max_grad_norm: 1.0
num_epochs: 1
optimizer: adamw_torch
output_dir: [REPLACE-WITH-MODEL-OUTPUT-PATH]
save_strategy: epoch
saves_per_epoch: 1
seed: 2442
sequence_length: 32768
sequence_parallel: true
torch_compile: false
trust_remote_code: true
use_tensorboard: true
val_set_size: 0.01
warmup_steps: 1000 # or remove for next-checkpoint
\end{lstlisting}
\end{PromptBox}

\subsection{Post-training Details}
\label{app:postraining-details}

Following the CPT phase, we performed post-training in two stages. 
First, we conducted supervised finetuning (SFT) using our augmented RedSage-Conv dataset together with general instruction data from the non-reasoning subset of SmolTalk2\footnote{General SFT datasets: \href{https://huggingface.co/HuggingFaceTB}{HuggingFaceTB/smoltalk2}}. 
This stage allowed the model to specialize in cybersecurity conversations while retaining general instruction-following capabilities. 

Second, we applied preference alignment via Direct Preference Optimization (DPO) using the open-source Tulu~3 8B Preference Mixture dataset~\citep{lambert2025tulu3pushingfrontiers}. 
This alignment phase refined the model's responses to better reflect human-preferred outputs.

The Axolotl configuration for the post-training stages is shown below:

\begin{PromptBox}[RedSage Supervised-Finetuning Config]
\begin{lstlisting}[basicstyle=\ttfamily\small,breaklines=true,showstringspaces=false]
base_model: [REPLACE-WITH-REDSAGE-BASE-MODEL]
trust_remote_code: true
auto_resume_from_checkpoints: true

bf16: true                                    
deepspeed: deepspeed_configs/zero3_bf16.json  
gradient_checkpointing: true                 
sequence_parallel: true                      

micro_batch_size: 32
gradient_accumulation_steps: 1
num_epochs: 2
sequence_length: 32768

optimizer: adamw_torch
lr_scheduler: cosine
learning_rate: 2.5e-5
weight_decay: 0.05
warmup_ratio: 0.01
cosine_min_lr_ratio: 0.01

chat_template: jinja
chat_template_jinja: [REPLACE-WITH-OUR-CUSTOM-CHAT-TEMPLATE]

datasets:
# Conversation Datasets
  - path: [REPLACE-WITH-REDSAGE-CONVERSATION-DATA]
    type: chat_template
    name: all
    field_messages: conversations
    message_property_mappings:
      role: from
      content: value

  - path: [REPLACE-WITH-SMOLTALK2-NON-THINKING]
    type: chat_template
    name: formatted
    field_messages: messages
    message_property_mappings:
      role: from
      content: value
output_dir: [REPLACE-WITH-MODEL-OUTPUT-PATH]
save_steps: 0.25
eval_steps: 0.25
val_set_size: 0.01

log_with:
  - wandb
  - tensorboard

use_tensorboard: true

save_total_limit: 5            
load_in_8bit: false            
torch_compile: false           

special_tokens:
  eos_token: <|im_end|>
  pad_token: <|endoftext|>
\end{lstlisting}
\end{PromptBox}

\subsection{\red{Estimated Training Time and Computational Cost Analysis}}
\label{app:compute-analysis}

\red{Continued pretraining from Qwen3-8B-Base on the CyberFineWeb (CFW) dataset was executed in 24-hour maximum-runtime chunks, with an average of 20 effective training hours per chunk. Five such chunks required approximately 100 hours to produce the RedSage-8B-CFW checkpoint. Additional continued pretraining on RedSage-Seed and RedSage-Dump took roughly 10 hours, yielding RedSage-8B-Base. Supervised fine-tuning on RedSage-Conv and general instruction datasets (SmolTalk2) required about 16 hours for two epochs, and DPO alignment using 8$\times$A100 GPUs added another 8 hours. In total, the full training pipeline consumed approximately 134 wall-clock hours ($\sim$5.5 days), corresponding to more than 4{,}000 GPU-hours. A detailed breakdown of each stage is provided in Table~\ref{tab:compute-analysis}. Variations may arise from distributed-training overheads, including communication latency and checkpoint restarts.}

\begin{table}[h]
\centering
\small
\caption{\red{Estimated training time and computational cost for the RedSage-8B pipeline.}}
\label{tab:compute-analysis}
\begin{tabularx}{\linewidth}{lXcc}
\toprule
\textbf{Stage} & \textbf{Output Checkpoint} & \textbf{Time (h)} & \textbf{GPU-hours (approx.)} \\
\midrule
\multicolumn{4}{l}{\textit{Continued Pretraining (CPT), 1 epoch, 32$\times$A100}} \\
\midrule
CPT: CyberFineWeb
& RedSage-8B-CFW
& $\sim$100
& $\sim$3{,}200 \\
CPT: RedSage-Seed \& -Dump
& RedSage-8B-Base
& $\sim$10
& $\sim$320 \\
\midrule
\multicolumn{4}{l}{\textit{Post-training (SFT: 2 epochs, 32$\times$A100; DPO: 1 epoch, 8$\times$A100)}} \\
\midrule
SFT: RedSage-Conv \& SmolTalk2
& RedSage-8B-Ins
& $\sim$16
& $\sim$512 \\
DPO: Tulu Preference Mixture
& RedSage-8B-DPO
& $\sim$8
& $\sim$64 \\
\midrule
\textbf{Total pipeline} 
& RedSage-8B-DPO
& \textbf{$\sim$134 ($\sim$5.5 days)}
& \textbf{$\sim$4{,}096} \\
\bottomrule
\end{tabularx}
\end{table}






\section{Evaluation Details}
\label{app:evaluation-details}

For replicable evaluation, we implement and evaluate RedSage-Bench and prior cybersecurity benchmarks in HuggingFace \texttt{lighteval}~\citep{lighteval}. The details of the compared models, tasks, and metrics for each evaluation are described in the next subsection.

\subsection{Evaluation Setup}
\label{app:evaluation-setups}

\paragraph{Compared methods.}
We benchmark RedSage against open general-purpose and cybersecurity-focused LLMs, summarized in Tab.~\ref{tab:baseline_hf_cards}. The general baselines are Llama-3.1-8B and Qwen3-8B; the specialized baselines are Llama-Primus (Base and Merged), Foundation-Sec (Base and Instruct), Lily-Cybersecurity-7B-v0.2, and DeepHat-V1-7B. For each model the table reports parameter count, backbone, and the Hugging Face card used to obtain configurations and weights, which supports strict reproducibility. Base models are evaluated in plain completion mode, instruction-tuned models use their official prompt templates, and Qwen3 is run in non-reasoning mode for parity. The suite spans 7-8B parameters across Llama, Qwen, and Mistral backbones, enabling a balanced comparison by capacity and training style.

\begin{table*}[h]
\centering
\caption{Evaluated baseline models and their Hugging Face cards.}
\label{tab:baseline_hf_cards}
\small
\setlength{\tabcolsep}{6pt}
\renewcommand{\arraystretch}{1.12}
\begin{tabularx}{\textwidth}{l c Y Y}
\toprule
\rowcolor{Gray}
\textbf{Model} & \textbf{Params (B)} & \textbf{Base model} & \textbf{Hugging Face} \\
\midrule
Llama-3.1-8B              & 8  & N/A (base)                                        & \href{https://huggingface.co/meta-llama/Llama-3.1-8B}{\url{meta-llama/Llama-3.1-8B}} \\
Qwen3-8B                  & 8  & Qwen3-8B-Base                                     & \href{https://huggingface.co/Qwen/Qwen3-8B}{\url{Qwen/Qwen3-8B}} \\
Llama-Primus-Base         & 8  & Llama-3.1-8B-Instruct                             & \href{https://huggingface.co/trend-cybertron/Llama-Primus-Base}{\url{trend-cybertron/Llama-Primus-Base}} \\
Llama-Primus-Merged       & 8 & Llama-3.1-8B (merged with Llama-3.1-8B-Instruct)  & \href{https://huggingface.co/trendmicro-ailab/Llama-Primus-Merged}{\url{trendmicro-ailab/Llama-Primus-Merged}} \\
Foundation-Sec-8B         & 8  & Llama-3.1-8B                                      & \href{https://huggingface.co/fdtn-ai/Foundation-Sec-8B}{\url{fdtn-ai/Foundation-Sec-8B}} \\
Foundation-Sec-8B-Instruct& 8. & Foundation-Sec-8B (Llama-3.1-8B backbone)         & \href{https://huggingface.co/fdtn-ai/Foundation-Sec-8B-Instruct}{\url{fdtn-ai/Foundation-Sec-8B-Instruct}} \\
Lily-Cybersecurity-7B-v0.2& 7 & Mistral-7B-Instruct-v0.2                          & \href{https://huggingface.co/segolilylabs/Lily-Cybersecurity-7B-v0.2}{\url{segolilylabs/Lily-Cybersecurity-7B-v0.2}} \\
DeepHat-V1-7B             & 7 & Qwen2.5-Coder-7B                                  & \href{https://huggingface.co/DeepHat/DeepHat-V1-7B}{\url{DeepHat/DeepHat-V1-7B}} \\
\bottomrule
\end{tabularx}
\end{table*}

\newpage

\subsection{RedSage Benchmarks}
\label{app:redsage-benchmarks}

\red{\textbf{MCQ Evaluation Protocols.} 
Models are prompted to select a single option letter (A-D) given a question and its choices. We compute the log probabilities of the option tokens for the next-token prediction and take the highest-probability option as the model's answer. This approach avoids parsing errors and ensures the model outputs only the option letter. The MCQ prompt template is shown below.}

\begin{PromptBox}[\red{RedSage MCQ Prompt Template}]
\begin{lstlisting}[breaklines=true,breakatwhitespace=true,columns=fullflexible,showstringspaces=false,basicstyle=\ttfamily\footnotesize]
You are given multiple choice questions. Answer with the option letter (A, B, C, D) from the given choices directly.
Question: {question}
A. {option[0]}
B. {option[1]}
C. {option[2]}
D. {option[3]}
Answer:
\end{lstlisting}
\end{PromptBox}

\red{\textbf{Open-ended Q\&A Evaluation Protocols.} We adopt an LLM-as-Judge rubric that assesses both factual correctness (True/False) and answer quality (0-10), considering helpfulness, relevance, depth, and level of detail. All judgments are produced using Llama-3.3-70B as the evaluator. The system prompt and template for the rubric are provided below.
}

\begin{PromptBox}[\red{RedSage Open-ended Q\&A LLM-as-Judge System Prompt}]
\begin{lstlisting}[breaklines=true,breakatwhitespace=true,columns=fullflexible,showstringspaces=false,basicstyle=\ttfamily\footnotesize]
You are the Judge for an Open-QA cybersecurity benchmark.

## Inputs
- question: the final user-facing prompt
- reference_answer: the ground-truth answer
- model_answer: the response produced by the evaluated model

## Goal
Evaluate model_answer on two aspects:
(1) factual correctness, and
(2) overall answer quality (0-10).

---

## Checklist

1) Correctness
- Is the answer factually accurate?
- Does it align with the reference_answer and grounded cybersecurity knowledge?
- Any hallucinations, contradictions, or false claims?

Output rule for correctness:
- True if correct and grounded
- False if factually incorrect, hallucinated, contradicted, or clearly wrong

2) Answer Quality Score (0-10)
Rate overall quality only if at least partially correct:
- Helpfulness (does it answer the question?)
- Relevance (focused and on-topic)
- Depth (reasoning or understanding)
- Level of detail (complete and specific enough)

Scoring guide:
- 10: perfect - accurate, complete, deep, fully relevant
- 8-9: strong - minor omissions or small inaccuracies
- 6-7: moderate - useful but lacking depth or detail
- 4-5: weak - vague, shallow, or incomplete
- 1-3: poor - limited usefulness or clarity
- 0: invalid or gibberish

---

## Instructions
- Use chain-of-thought privately, but present only a final analysis in <analysis>.  
- Be strict on correctness: any factual error -> correctness=False. If correctness=False, cap score at 3 or lower.  
- If correct but shallow, keep correctness=True and assign a lower score.

---

## Output Format
Return exactly these three blocks in order. Do not add text outside the tags.

<analysis>
Free-form justification. You may write anything here such as step-by-step reasoning, comparisons, errors spotted, strengths, weaknesses, etc. between the model_answer and reference_answe.
Make sure your analysis is detailed and covers all aspects of the evaluation checklist.

### Correctness
Analysis and justification for the correctness evaluation.

### Answer Quality Score
Analysis and justification for the answer quality score.

#### Helpfulness
Justification for the helpfulness aspect.

#### Relevance
Justification for the relevance aspect.

#### Depth
Justification for the depth aspect.

#### Level of Detail
Justification for the level of detail aspect.
</analysis>

<correctness>
True or False
</correctness>

<score>
0-10 (integer only)
</score>
\end{lstlisting}
\end{PromptBox}

\begin{PromptBox}[\red{RedSage Open-ended Q\&A LLM as Judge Prompt Template}]
\begin{lstlisting}[breaklines=true,breakatwhitespace=true,columns=fullflexible,showstringspaces=false,basicstyle=\ttfamily\footnotesize]
[System Prompt]
Question:
```
{question}
```

Reference Answer:
```
{reference_answer}
```

Model Answer:
```
{model_answer}
```
\end{lstlisting}
\end{PromptBox}

\begin{figure}[h]
    \centering
    \includegraphics[width=1\linewidth]{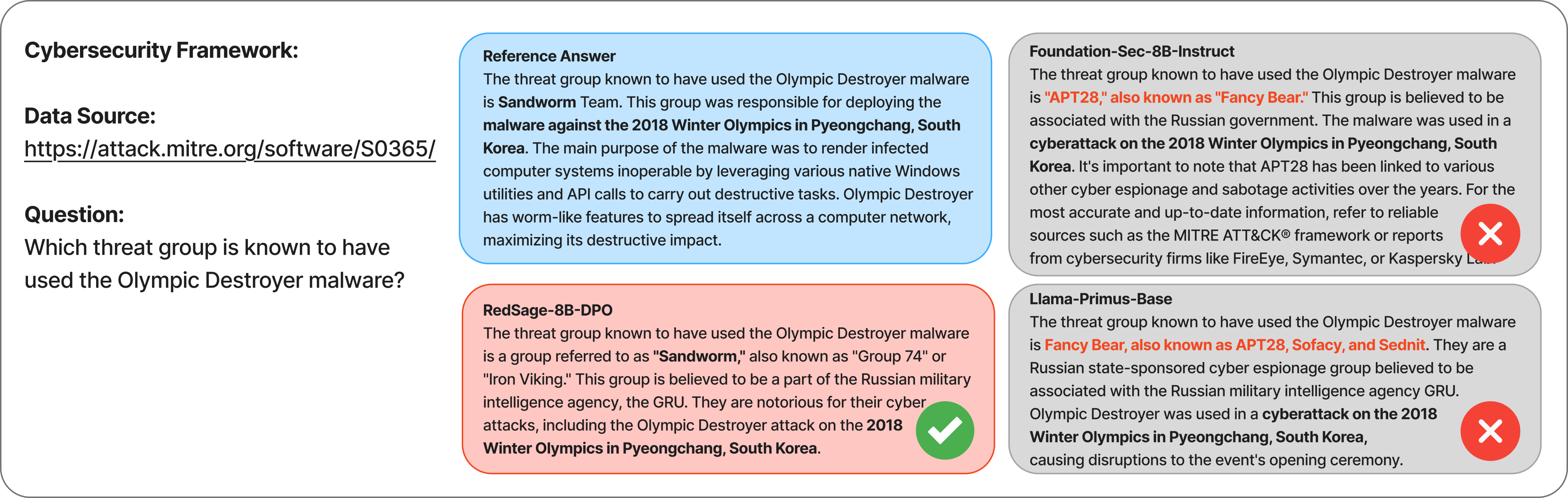}
    \caption{Knowledge framework example from RedSage OpenQA. For the Olympic Destroyer question, RedSage 8B DPO matches the reference attribution to the Sandworm team, while baseline models misattribute it to different APT groups. \textit{Best viewed in Zoom}.}
    \label{fig:openqa_framework}
\end{figure}

\begin{figure}[h]
    \centering
    \includegraphics[width=1\linewidth]{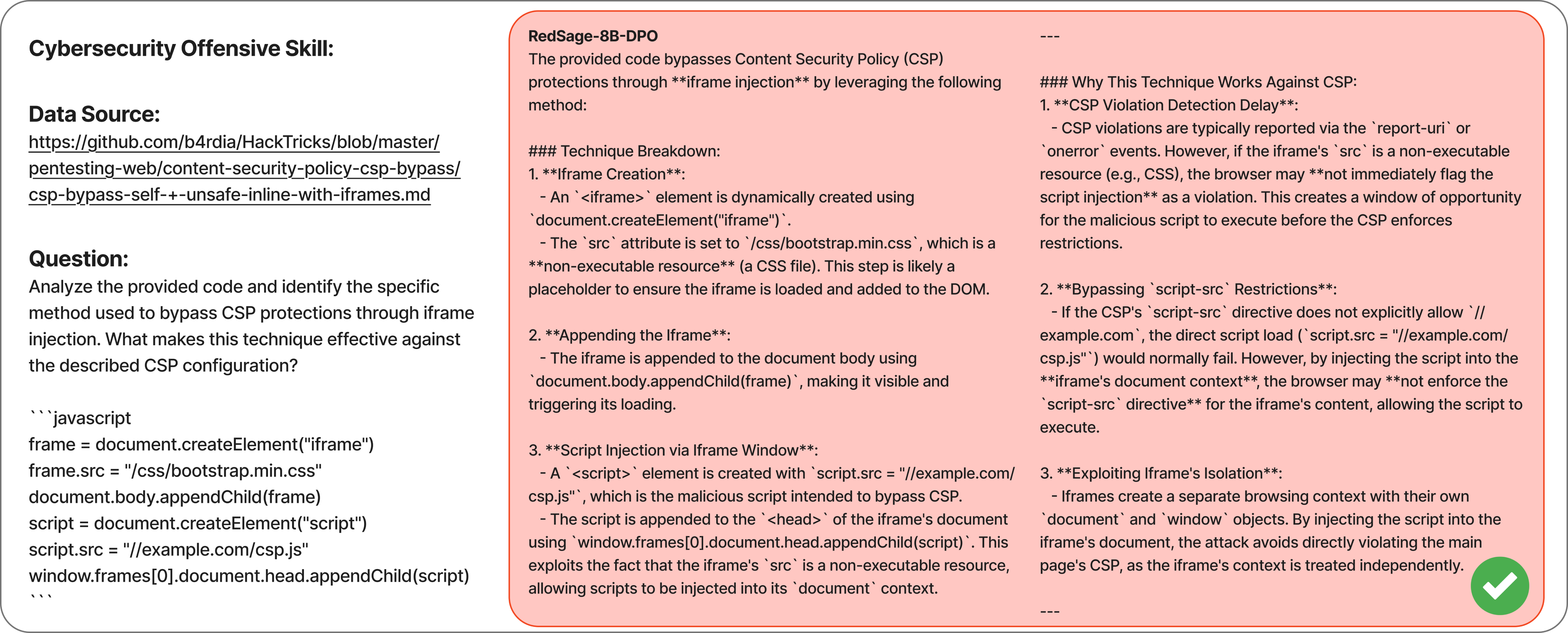}
    \caption{Offensive skill example analyzing JavaScript that bypasses Content Security Policy. RedSage 8B DPO correctly identifies iframe injection as the evasion technique and explains how each step circumvents the configured \texttt{script-src} directive. \textit{Best viewed in Zoom}.}
    \label{fig:openqa_offensive_skill}
\end{figure}

\begin{figure}[h]
    \centering
    \includegraphics[width=1\linewidth]{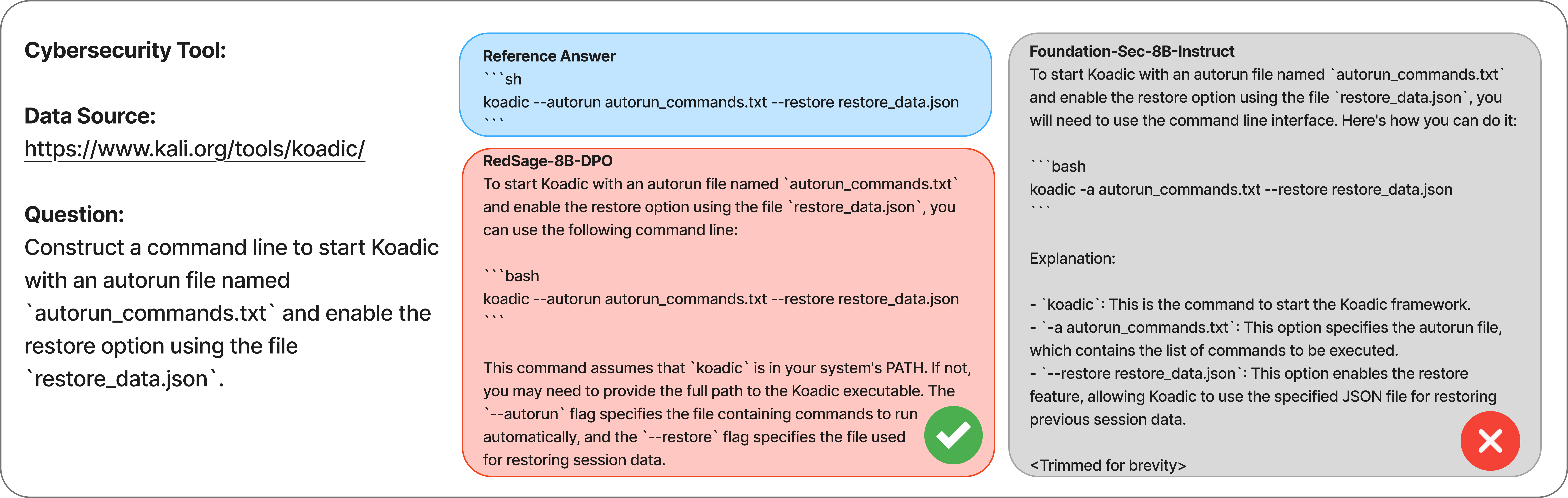}
    \caption{Tool-usage example for the Koadic framework. RedSage 8B DPO provides the exact command, while the baseline model produces a non-matching command. \textit{Best viewed in Zoom}.}
    \label{fig:openqa_tool}
\end{figure}

\textbf{Qualitative Results of RedSage OpenQA.}
We present three RedSage OpenQA examples that span cybersecurity frameworks, offensive skills, and tool usage. In the Olympic Destroyer attribution case shown in Fig.~\ref{fig:openqa_framework}, RedSage 8B DPO correctly identifies the Sandworm team, while baseline models misattribute the malware to other Russian APT groups. For the CSP bypass example in Fig.~\ref{fig:openqa_offensive_skill} and the Koadic tool-usage example in Fig.~\ref{fig:openqa_tool}, RedSage 8B DPO accurately recognizes iframe-based CSP evasion and generates the exact Koadic command line. In contrast, the baselines omit key details or produce malformed commands. These cases illustrate that RedSage exhibits stronger grounding in authoritative cybersecurity sources and improved precision in operational reasoning.

\textbf{Qualitative Results of LLM-as-Judge.}
To further illustrate the differences captured by our LLM-as-Judge pipeline, we include qualitative evaluations comparing RedSage with the baseline model using the tool-based question shown in Fig.~\ref{fig:openqa_tool}. As shown in Fig.~\ref{fig:llm-judge-qual}, the judge marks RedSage's answer as fully correct, assigns a perfect score, and highlights the precise command construction and clear supporting explanations. In contrast, the baseline model receives a failing correctness label and a substantially lower score because it uses an incorrect command-line flag, even though its surrounding explanation is detailed. These paired results emphasize the sensitivity of our evaluation framework to fine-grained correctness, particularly in cybersecurity scenarios where small syntactic deviations can lead to incorrect or unsafe tool behavior.


\begin{tcolorbox}[title=RedSage: LLM-as-Judge Output, breakable, colback=white]
\footnotesize
\begin{lstlisting}[
    breaklines=true,
    breakatwhitespace=true,
    columns=fullflexible,
    showstringspaces=false,
    basicstyle=\ttfamily\footnotesize,
    breakindent=0pt,       % Sets the wrap indent to zero
    breakautoindent=false  % Prevents auto-indentation based on the previous line
]
<analysis>
The model answer provided is not only factually correct but also offers additional helpful information and context. It accurately constructs the command line to start Koadic with the specified autorun file and restore option. The command line provided in the model answer matches the reference answer exactly, which is `koadic --autorun autorun_commands.txt --restore restore_data.json`. This indicates that the model has correctly understood the requirements and has provided the precise command needed to achieve the desired outcome.

### Correctness
The model answer is correct because it accurately reflects the command line needed to start Koadic with the specified parameters. There are no factual errors, hallucinations, or contradictions in the provided command line. The additional information about the command assuming `koadic` is in the system's PATH and the explanation of the flags (`--autorun` and `--restore`) are accurate and useful but do not affect the correctness of the command itself.

### Answer Quality Score
The answer quality score should be high because the model answer is not only correct but also provides additional useful information that could help a user understand how to use the command and troubleshoot potential issues (like the executable not being in the PATH).

#### Helpfulness
The model answer is very helpful because it directly answers the question with the correct command line and provides supplementary information that could aid in successfully executing the command.

#### Relevance
The answer is highly relevant as it directly addresses the question asked without including unnecessary information. The command line and the explanations provided are focused on the task of starting Koadic with an autorun file and enabling the restore option.

#### Depth
The answer shows a good depth of understanding of the command and its parameters. It not only provides the command but also explains the purpose of the flags used, which indicates a level of comprehension beyond mere recall of command syntax.

#### Level of Detail
The level of detail is appropriate for the question asked. The model provides enough detail for someone to understand how to use the command and what each part of the command does, without overwhelming them with unnecessary information.

</analysis>

<correctness>
True
</correctness>

<score>
10
</score>
\end{lstlisting}
\end{tcolorbox}


\begin{tcolorbox}[title=Baseline Model: LLM-as-Judge Output, breakable, colback=white]
\footnotesize
\begin{lstlisting}[
    breaklines=true,
    breakatwhitespace=true,
    columns=fullflexible,
    showstringspaces=false,
    basicstyle=\ttfamily\footnotesize,
    breakindent=0pt,       % Sets the wrap indent to zero
    breakautoindent=false  % Prevents auto-indentation based on the previous line
]
<analysis>
The model answer provides a detailed explanation of how to start Koadic with an autorun file and enable the restore option. It breaks down the command into its components and explains the purpose of each part. However, the command provided in the model answer is slightly different from the reference answer. The model answer uses the short option `-a` for autorun, whereas the reference answer uses the long option `--autorun`. 

### Correctness
The model answer is almost correct, but it uses a different option for specifying the autorun file. The correct option, as per the reference answer, is `--autorun`, not `-a`. This discrepancy makes the model answer factually incorrect.

### Answer Quality Score
Despite the factual error, the model answer provides a clear and detailed explanation of the command and its components. It also includes important notes about the ethical use of Koadic and the importance of proper permissions and access.

#### Helpfulness
The model answer is helpful in explaining the purpose of each part of the command and providing context for its use.

#### Relevance
The model answer is highly relevant to the question, as it directly addresses the task of starting Koadic with an autorun file and enabling the restore option.

#### Depth
The model answer provides a good depth of explanation, breaking down the command into its components and explaining the purpose of each part.

#### Level of Detail
The model answer provides a sufficient level of detail, including explanations of the command options and important notes about ethical use and permissions.
</analysis>

<correctness>
False
</correctness>

<score>
3
</score>
\end{lstlisting}
\end{tcolorbox}

\captionof{figure}{Qualitative LLM-as-Judge outputs comparing RedSage and the baseline model.}
\label{fig:llm-judge-qual}

\newpage

\subsection{Cybersecurity Benchmarks}
\label{app:cybersecurity-benchmarks}

\paragraph{CyberMetric (CyMtc).}
CyberMetric evaluates general cybersecurity knowledge via multiple-choice questions with four options, curated from authoritative sources such as NIST publications, RFCs, books, and research papers using a retrieval-augmented generation pipeline. The collection is released in several sizes, and we use the 500-item split that was fully verified by human experts. Items span nine topical areas that include cryptography, reverse engineering, and risk assessment. Models are scored with standard MCQ accuracy. \citep{cybermetric24}

\paragraph{SecBench (ScBen).}
SecBench is a large multi-dimensional benchmark for cybersecurity that includes both MCQs and short-answer questions, covers two capability levels (knowledge retention and logical reasoning), and is available in Chinese and English. Questions were sourced from open materials and a curated contest, and short-answer evaluation is supported by an LLM-based grader. In our study we use the English MCQ subset and report accuracy. \citep{jing2024secbench}

\paragraph{MMLU Computer Security (MMLU-CSec).}
MMLU is a 57-subject multiple-choice test that measures broad academic and professional knowledge. We evaluate on the Computer Security subject, which contains MCQs covering practical and theoretical topics such as network security and cryptography. Following common practice for MMLU-style evaluation, we report accuracy. \citep{hendryckstest2021}

\paragraph{SECURE.}
SECURE targets applied cybersecurity with datasets built from MITRE ATT\&CK, CWE, CVE, and related ICS advisories, organized into three knowledge types: extraction, understanding, and reasoning. We use the MCQ-style subsets MAET (MITRE ATT\&CK Extraction), CWET (Common Weakness Extraction), and KCV (Knowledge test on Common Vulnerabilities). The authors manually refined the pools by removing or fixing flawed questions. We evaluate with MCQ accuracy. \citep{secure2024}

\paragraph{CTI-Bench.}
CTI-Bench focuses on cyber threat intelligence and provides four tasks: CTI-MCQ for knowledge of CTI standards and practices; CTI-RCM for mapping CVE descriptions to one or more CWE root causes; CTI-VSP for predicting CVSS v3 base vectors and scores; and CTI-ATE for extracting MITRE ATT\&CK attack techniques from natural language incident descriptions. While VSP and ATE are typically evaluated with regression and F1 metrics, respectively, in our study we only use accuracy across all subsets for consistent aggregation. \citep{ctibench2024}

\paragraph{SecEval (ScEva).}
SecEval is a domain-focused benchmark of more than two thousand MCQs spanning nine areas that include software, application, system, web, cryptography, memory safety, network security, and penetration testing. Questions were constructed from textbooks, official documentation, and standards using GPT-4 prompting, with quality control to remove invalid items. We evaluate with MCQ accuracy on the full set. \citep{li2023seceval}

\subsection{General LLM Benchmarks}
\label{app:general-benchmarks}

\paragraph{ARC-Challenge (ARC-C).}
ARC-C is the challenge split of the AI2 Reasoning Challenge, a set of grade-school science multiple-choice questions curated to require nontrivial reasoning and background knowledge. The challenge subset specifically contains items that defeat simple retrieval and co-occurrence baselines, making it a strong discriminator of reasoning beyond surface cues. We evaluate with standard MCQ accuracy as used by leaderboard implementations. \citep{Clark2018ThinkYH}

\paragraph{HellaSwag (HSwag).}
HellaSwag tests grounded commonsense inference via sentence completion. Each example presents a short context and four candidate endings that describe plausible next events in physical or social scenarios. The dataset was adversarially filtered to foil strong language models while remaining trivial for humans, which sharpens its discriminative power. Performance is reported as multiple-choice accuracy. \citep{zellers2019hellaswag}

\paragraph{TruthfulQA (TQA).}
TruthfulQA measures whether models avoid widespread misconceptions and misleading patterns by answering with factually truthful content across 38 categories such as health, law, and finance. It provides both generative prompts and multiple-choice variants. Following common leaderboard practice, we use the multiple-choice setting and report accuracy to ensure comparability across models. \citep{lin2021truthfulqa}

\paragraph{MMLU.}
MMLU evaluates broad knowledge and reasoning across 57 academic and professional subjects that range from elementary mathematics and U.S. history to computer science and law. Each subject consists of four-option multiple-choice items designed to test recall, conceptual understanding, and problem solving. Scores are aggregated as average accuracy across subjects. \citep{hendryckstest2021}

\paragraph{WinoGrande (WinoG).}
WinoGrande is a large adversarial variant of the Winograd Schema Challenge that assesses commonsense reasoning through pronoun resolution. Each example requires selecting which of two candidate nouns a pronoun refers to, with items constructed to reduce annotation artifacts and shallow heuristics. Evaluation follows leaderboard protocol using accuracy. \citep{sakaguchi2020winogrande}

\paragraph{GSM8K.}
GSM8K is a collection of 8.5K carefully authored grade-school math word problems that require multi-step arithmetic reasoning. Problems are linguistically diverse and designed to encourage chain-of-thought solutions, yet the final target is a short numeric answer. We report exact-match accuracy on the final answer, consistent with leaderboard settings. \citep{cobbe2021gsm8k}

\paragraph{IFEval.}
IFEval evaluates instruction following using prompts that contain verifiable constraints such as minimum length, required keywords, or structural requirements. Each prompt includes one or more constraints that can be programmatically checked, yielding objective pass/fail signals without human grading. We report the mean compliance rate across all constraints, i.e., the percentage of constraints satisfied. \citep{zhou2023ifeval}

\section{\red{Additional Evaluation Results}}

\subsection{\red{Larger Model Scaling}}
\label{app:larger-scaling}

\red{To assess the scalability of our data curation and augmentation pipeline, we conducted a reduced-scope experiment using Qwen3-32B. We applied QLoRA fine-tuning ($\approx$1\% trainable parameters) on a partial dataset consisting of the curated RedSage-Seed subset (excluding RedSage-Dump) and 50\% of RedSage-Conv. Despite using only a fraction of the full training data and a lightweight adaptation method, the resulting 32B model achieved consistent gains across both the RedSage-MCQ benchmark (Table~\ref{tab:redsage-mcq-scale}) and a suite of cybersecurity evaluations (Table~\ref{tab:cyber-bench-scale}). Notably, the training loss continued to decrease throughout the run, suggesting that full-data, full-parameter fine-tuning would yield even larger improvements. These findings indicate that the RedSage data curation and augmentation methodology transfers effectively to larger models, underscoring its scalability and potential to advance cybersecurity LLM development.}

\begin{figure*}[h]
\centering \small
\centering
\captionof{table}{\red{RedSage-MCQ (0-shot) scaling experiment. Values are accuracy (\%). 
Abb: Gen = General, Frm = Frameworks, 
Off = Offensive Skills, CLI = Command-line Tools, Kali = Kali Tools.}}
\label{tab:redsage-mcq-scale}
\begin{tabular}{l c c c c c c}
\toprule
\rowcolor{Gray} 
\textbf{Model Name} & 
\multicolumn{1}{c}{\textbf{Macro}} & 
\multicolumn{2}{c}{\textbf{Knowledge}} & 
\multicolumn{1}{c}{\textbf{Skill}} & 
\multicolumn{2}{c}{\textbf{Tools}} \\
\cmidrule(lr){3-4} \cmidrule(lr){5-5} \cmidrule(lr){6-7}
 & \textbf{Acc} & \textbf{Gen} & \textbf{Frm} & \textbf{Off} & \textbf{CLI} & \textbf{Kali} \\
\midrule
Qwen3-8B              & 81.85 & 80.46 & 78.82 & 86.16 & 83.92 & 75.56 \\
Qwen3-32B             & 85.40 & 84.08 & 82.32 & 89.00 & 87.60 & 80.40 \\
RedSage-8B-Ins        & 85.73 & 84.20 & 84.98 & 89.06 & 86.80 & 80.30 \\
RedSage-32B-LoRA-Ins-0.5   & \textbf{87.53} & \textbf{85.68} & \textbf{85.04} & \textbf{91.46} & \textbf{88.76} & \textbf{82.78} \\
\bottomrule
\end{tabular}
\end{figure*}

\begin{table*}[h]
\centering \small

\caption{\red{Related Cybersecurity Benchmarks (0-shot) scaling experiment. 
Values are Accuracy (\%). Best results are shown in bold.}}
\label{tab:cyber-bench-scale}
\setlength{\tabcolsep}{5.5pt}
\scalebox{0.85}[0.85]{
\begin{tabular}{lcccccccccc}
\toprule
\rowcolor{Gray}
\multirow{2}{*}{\textbf{Model Name}} & \multirow{2}{*}{\textbf{Mean}} & \multicolumn{2}{c}{\textbf{CTI-Bench}} & \multicolumn{1}{c}{\textbf{CyMtc}} & \multicolumn{1}{c}{\textbf{MMLU}} & \multicolumn{1}{c}{\textbf{ScBen}} & \multicolumn{1}{c}{\textbf{ScEva}} & \multicolumn{3}{c}{\textbf{SECURE}} \\
\cmidrule(lr){3-4} \cmidrule(lr){5-5} \cmidrule(lr){6-6} \cmidrule(lr){7-7} \cmidrule(lr){8-8} \cmidrule(lr){9-11}
& & \textbf{MCQ} & \textbf{RCM} & \textbf{500} & \textbf{CSec} & \textbf{En} & \textbf{MCQ} & \textbf{CWET} & \textbf{KCV} & \textbf{MAET} \\
\midrule
Qwen3-8B            & 75.71 & 62.76 & 54.00 & 88.60 & 76.00 & 73.26 & 65.46 & 88.11 & 87.42 & 85.75 \\
Qwen3-32B           & 82.31 & 70.04 & 65.60 & 91.80 & \textbf{84.00} & \textbf{84.23} & 76.23 & 89.46 & \textbf{88.72} & 90.06 \\
RedSage-8B-Ins      & 81.30 & 70.56 & \textbf{76.70} & 89.80 & 78.00 & 79.91 & 72.48 & 91.45 & 81.34 & 91.47 \\
RedSage-32B-LoRA-Ins-0.5 & \textbf{82.85} & \textbf{71.64} & 66.10 & \textbf{93.40} & \textbf{84.00} & 83.77 & \textbf{78.30} & \textbf{92.18} & 83.29 & \textbf{92.97} \\
\bottomrule
\end{tabular}
}
\end{table*}

\end{document}